\documentclass[prb,aps,amssym,nofootinbib,floatfix,eqsecnum,twocolumn,notitlepage]{revtex4-1} 
\usepackage{xcolor}
\usepackage{amsmath,amssymb,bbold,bm}
\usepackage{graphicx}
\usepackage{mathdots}
\usepackage{caption}
\usepackage{subcaption}
\usepackage{cancel}
\usepackage{comment}
\usepackage{currfile}

\newcommand{\br}{\nonumber\\*}
\newcommand{\be}{\begin{equation}}
\newcommand{\ben}{\begin{equation*}}
\newcommand{\ee}{\end{equation}}
\newcommand{\een}{\end{equation*}}
\newcommand{\bs}{\begin{split}}
\newcommand{\es}{\end{split}}
\newcommand{\bmx}{\begin{array}}
\newcommand{\emx}{\end{array}}
\newcommand{\bea}{\begin{eqnarray}}
\newcommand{\bean}{\begin{eqnarray*}}
\newcommand{\eea}{\end{eqnarray}}
\newcommand{\eean}{\end{eqnarray*}}
\newcommand{\dg}{^{\dagger}}
\newcommand{\dn}{^{\vphantom{\dagger}}}

\newcommand{\vsp}{\vphantom{\Big]}}
\newcommand{\lr}{\leftrightarrow}
\newcommand{\ra}{\rightarrow}

\newcommand{\ua}{\uparrow}
\newcommand{\da}{\downarrow}
\newcommand{\qqquad}{\qquad\qquad\qquad}

\newcommand{\eps}{\epsilon}
\newcommand{\pref}[1]{(\ref{#1})}

\newcommand{\intinf}[1]{\int_{-\infty}^{+\infty}{#1}}
\newcommand{\intoinf}[1]{\int_{0}^{\infty}{#1}}
\newcommand{\intopi}[1]{\int_{0}^{\pi}{#1}}

\newcommand{\intob}[1]{\int_{0}^{\beta}{#1}}

\newcommand{\re}[1]{{\rm Re}\left[ #1 \right]}
\newcommand{\im}[1]{{\rm Im}\left[ #1 \right]}
\newcommand{\tr}[1]{{\rm Tr}\Big[ #1 \Big]}

\newcommand{\bw}[1]{\begin{widetext}}
\newcommand{\ew}[1]{\end{widetext}}

\newcommand{\abs}[1]{\left\vert #1 \right\vert}
\newcommand{\bra}[1]{\left\langle #1 \right\vert}
\newcommand{\ket}[1]{\left\vert #1\right\rangle}
\newcommand{\braket}[1]{\left\langle #1\right\rangle}
\newcommand{\brket}[1]{\langle #1\rangle}

\newcommand{\com}[2]{\left[#1,#2\right]}
\newcommand{\acom}[2]{\left\{#1,#2\right\}}
\newcommand{\mat}[1]{\left(\bmx{cc}#1\emx\right)}
\newcommand{\matn}[1]{\bmx{cc}#1\emx}
\newcommand{\matl}[1]{\bmx{ll}#1\emx}

\newcommand{\nq}{\mathsf}
\newcommand{\bnq}[1]{\breve{\mathsf #1}}
\newcommand{\bv}[1]{\breve{#1}}

\begin{document}
\title{Interaction effects in Aharonov-Bohm-Kondo Rings}
\author{Yashar Komijani$^1$}\email{komijani@phas.ubc.ca}
\author{Ryosuke Yoshii$^{2,3}$}
\author{Ian Affleck$^1$}

 \affiliation{$^1$Department of Physics and Astronomy and Quantum Materials Institute, University of British
Columbia, Vancouver, B.C., Canada, V6T 1Z1\\
$^2$Yukawa Institute for Theoretical Physics, Kyoto University,
Kitashirakawa Oiwake-Cho, Kyoto 606-8502, Japan\\
$^3$Research and Education Center for Natural Sciences,
Keio University, 4-1-1 Hiyoshi, Kanagawa 223-8521, Japan
}

\begin{abstract}
We study the conductance through an Aharonov-Bohm ring, containing a quantum dot in the Kondo regime in one arm, 
at finite temperature and arbitrary electronic density.  We develop a general method 
for this calculation based on changing basis to the screening and non-screening channels. We show that 
an unusual term appears in the conductance, involving the connected 4-point Green's function of the conduction electrons.
However, this term and terms quadratic in the ${\rm T}$-matrix can be eliminated at sufficiently low temperatures, leading to an expression for the conductance
linear in the Kondo T-matrix. 
Explicit results are given for temperatures high compared to the Kondo temperature.
\end{abstract}
\date{\today}
\maketitle
\section{Introduction}
The Kondo effect [\onlinecite{Kondo},\onlinecite{Hewson}] is the entanglement of an impurity spin with conduction electrons as the temperature is lowered below 
the Kondo temperature $T_K$ and the effective Kondo coupling renormalizes to large values. The Kondo effect continues 
to fascinate, especially since its experimental realization in gated semiconductor quantum dots [\onlinecite{GG},\onlinecite{Cronenwett}]. One interesting 
and highly non-trivial extension of the basic Kondo model involves an Aharonov-Bohm (AB) ring with a Kondo impurity in 
the upper arm and interference with a lower ``reference arm'' which we refer to as an Aharonov-Bohm-Kondo (ABK) ring. Early works on this topic [\onlinecite{Bruder}-\onlinecite{Yoshii}] were largely focussed on the $T=0$ 
behaviour, using Nozi\`eres local Fermi liquid theory (FLT) [\onlinecite{Nozieres}] and Numerical Renormalization Group (NRG) [\onlinecite{NRG}] techniques. A conclusion of 
these works
was that $T_K$ depends strongly on the magnetic flux through the ring. As shown in 
[\onlinecite{Zarand04}-\onlinecite{Micklitz06}] Kondo scattering has both elastic and inelastic components which exhibit interesting 
variations with energy scale. Since inelastic scattering is generally known to destroy interference effects, 
a correspondingly rich dependence of the conductance of an ABK ring on temperature and flux is expected. Recently, the finite temperature behaviour of the ABK ring was also considered [\onlinecite{Yoshii,Carmi}]. [\onlinecite{Carmi}] studied the role of inelastic scattering on the visibility of AB oscillations through a large open ring with an embedded quantum dot. 
Various assumptions were made in this work including the idealized notion of a large `open ring' which relate the conductance to the scattering cross section. Here we avoid these assumptions, calculating the full conductance including 
contributions from multiple traversals of the ring,  using the Kubo formula.  

\begin{figure}[h!]
\centering
\includegraphics[scale=0.34]{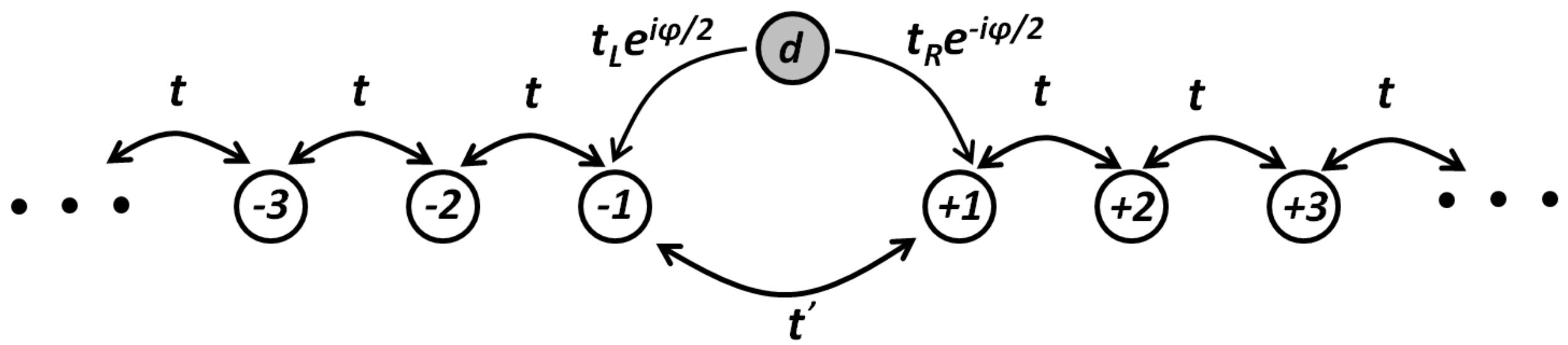}
\caption{\raggedright\small The model of a small ABK ring considered in this paper. The reference arm is a direct link between sites -1 and 1. The quantum dot, denoted by $d$, consists of a single level, with an on-site Coulomb repulsion, tunnel-coupled to the ring. The ring encloses a flux $h\varphi/e$ which is included symmetrically in the phases of the tunneling amplitudes of the dot in the gauge chosen here.}
\label{fig:Fig1}
\end{figure}

We  consider the short ABK ring introduced in [\onlinecite{Hofstetter}] with tunneling amplitudes $t_L$ and $t_R$ 
between a quantum dot and left and right leads along with a direct tunneling amplitude $t'$ between the leads. 
[See Fig.\,(\ref{fig:Fig1}).]
Unlike most previous work focusing on zero temperature, we study the temperature regime $T\gg T_K$. 
In this regime we expect renormalization group improved perturbation theory in the Kondo coupling, $J$, to be valid. Following earlier works [\onlinecite{Kondoscattering}, \onlinecite{Bruder}, \onlinecite{MA},  \onlinecite{Yoshii}] we develop a general method to study this system based on 
changing to a basis of channels of conduction electrons, $(\psi ,\phi)$ where $\psi$ interacts with the quantum dot 
and $\phi$ does not. This involves first transforming to the scattering state basis in the case $t_L=t_R=0$ and 
the transmission is only through the reference arm. Although the resulting Kondo coupling to $\psi$ is a complicated 
function of all parameters in the model, including the flux, this transformation has the advantage that one can then 
invoke known universal results on the standard single channel Kondo model. In particular, [\onlinecite{Hofstetter}] 
stated a formula for the conductance where all interaction effects were contained in the single-electron ${\rm T}$-matrix 
for the corresponding single-channel Kondo model.
The frequency and temperature dependence of 
${\rm T}(\omega ,T)$ has been well studied using renormalization group improved perturbation theory, FLT, NRG 
and other methods.  Starting from the Kubo formula, we show that the conductance can be written in terms of a `transmission probability' function which has
a disconnected two-point function part (of zeroth, first and second order in the T-matrix) and a 
connected four-point function part $\mathcal{G}^C=\braket{\psi^\dagger \psi \psi^\dagger \psi}_C$. However, using a suitable formulation 
of the Kubo formula, it is possible to eliminate both the connected term and the term quadratic in the  ${\rm T}$-matrix at temperatures small compared to the 
band width, which could still be large compared to $T_K$, resulting in an expression for the conductance 
linear in the T-matrix. We relate this result to results of Meir and Wingreen [\onlinecite{Meir}, \onlinecite{Meir2}] showing that the conductance through a rather general 
interacting central region can be expressed in terms of the ${\rm T}$-matrix. Similar formulas have been obtained in the past using Keldysh technique [\onlinecite{Koenig}-\onlinecite{Hofstetter}]. Here we rederive them using Kubo formalism, 
generalize them to finite temperature and arbitrary density in the leads and examine critically when they are valid.

In Sec.\,\pref{sec:model} we introduce our model in more detail and show how we can transform the ABK ring model to a
single channel Anderson or  Kondo model. In Sec.\,\pref{sec:cond} the Kubo formula for the 
linear conductance is expressed in the $(\psi ,\phi)$ basis, containing terms involving the ${\rm T}$-matrix 
as well as $\mathcal{G}^C$ and we present our formula for the conductance in terms of the T-matrix and connected Green's function. 
In Sec.\,\pref{sec:pert} we perform 
a perturbative calculation of both connected and disconnected parts to second order in the Kondo coupling and give 
an explicit formula and graphs for the flux-dependent conductance at high temperature $T\gg T_K$. In Sec.\,\pref{sec:con} we discuss the exact 
or approximate elimination of the connected Green's function from the conductance, using both Kubo and Keldysh formalisms and explain why the Meir-Wingreen approach does not generally allow for an exact elimination of the connected part although it does for some simpler models including special cases of the ABK ring.
We also present a formula, Eq.\,(\ref{G}),  for the conductance, containing only terms of zeroth and first order in the ${\rm T}$-matrix, which 
should be valid at temperatures small compared to the band width. 
Sec.\,\pref{conc} contains 
our conclusions and a discussion of open questions. 
Appendices \pref{sec:ppi} and \pref{sec:ap02} provide details related to the conductance calculations in the paper. 
In Appendix \pref{sec:non-int}
the non-interacting limit of the ABK ring is discussed  using Landauer, Fisher-Lee [\onlinecite{FisherLee}] and Keldysh techniques and the role of inelastic scattering is commented on. 

\section{The models: Screening and non-Screening Channels}
\label{sec:model}
In this paper we consider interaction effects in a small Aharanov-Bohm ring with a quantum dot embedded in the upper arm. This system is modeled by a tight-binding Hamiltonian in which a direct link between first sites of left and right chains plays the role of the reference arm. Admittedly, transport experiments are usually performed on rings that are much larger than the one considered here, but as many references have studied this model, we choose to discuss it in order to illustrate various interesting features of the calculation. Moreover, the advantage of this model is that a direct solution of the non-interacting case is relatively simple and provides us with the possibility to confirm our Kubo calculations with various cross checks using both Keldysh and Landauer techniques. Once the method is established, generalization to larger rings and/or continuum models is straightforward.

\noindent 
The Hamiltonian is given by $H=H_0+H_{T}+H_d$ which consist of a non-interacting part
\be\label{eqH1}
H_0=-t\Bigg[\Bigg(\sum_{n=-\infty}^{-2}+\sum_{n=1}^{\infty}\Bigg) c\dg_nc\dn_{n+1}+h.c.\Bigg]-t'(c\dg_{-1}c\dn_1+h.c.)
\ee
composed of two semi-infinite leads, with the hopping parameter $t$,  coupled together with an amplitude $t'$ which plays the role of the reference arm. A sum over spin indices is implied. The only interacting part of the model involves the quantum dot described by
\be\label{eqH2}
H_d=\eps_dd\dg d+Un_{d\ua}n_{d\da}.
\ee
Here $n_{d\sigma}=d\dg_{\sigma}d\dn_{\sigma}$ for $\sigma=\ua,\da$. The quantum dot is connected to the leads via flux-dependent tunnel couplings
\be\label{eqH3}
H_T=-\Big[\Big(t_Le^{i\varphi/2}c\dg_{-1}+t_Re^{-i\varphi/2}c\dg_1\Big)d+h.c.\Big].
\ee
In the absence of the dot and the reference arm, the electron wave-functions at $n=\pm 1$ are of the form $\sin(ka)$ where $a$ is the lattice spacing, resulting in an energy-dependence of the tunneling amplitudes. (Some previous works~[\onlinecite{Hofstetter},\onlinecite{Yoshii}] have neglected this energy-dependence in order to simplify the calculations). Unless explicitly mentioned, we assume $a= 1$ and drop it from the formulas.

We are interested in the case $t_L^2$, $t_R^2\ll Ut$, so that the dimensionless Kondo coupling $\nu J~\propto (t_L^2+t_R^2)/(tU)\ll 1$ and universal behaviour 
characteristic of the Kondo effect may occur. On the other hand, we {\it do not} assume $t'$ is small compared to $t$ since this may not be 
the case in experiments and is not necessary to see manifestations of the Kondo effect.

\subsection{Screening and non-screening channels}\label{sec:screening}
Transformation of the ABK ring model to a single-impurity Anderson model using scattering states has been introduced in [\onlinecite{Kondoscattering}, \onlinecite{Bruder}, \onlinecite{MA}, \onlinecite{Yoshii}]. The Hamiltonian without the dot can be diagonalized with a pair of degenerate scattering states $H_0\ket{e_{k}}=\eps_k\ket{e_{k}}$ and $H_0\ket{o_{k}}=\eps_k\ket{o_{k}}$ for $k\in(0,\pi)$ which span the Hilbert space and are chosen to be orthogonal. $H_0$ has parity symmetry and these states are even and odd eigenstates of the parity operator, i.e. their wave-functions satisfy $\chi_{ek}(-n)=\chi_{ek}(n)$ and $\chi_{ok}(-n)=-\chi_{ok}(n)$ and for $n>0$ can be written as
\be\label{eqasymchi}
\chi_{(e/o)k}(n)=2e^{i\delta_k^\pm}\sin(nk+\delta_k^\pm).
\ee
For the present problem, the phase shifts have been calculated in [\onlinecite{MA}] and satisfy the relation
\be\label{eqsmallabkphase}
\sin\delta^{\pm}_k=\pm\tau\sin(k+\delta^{\pm}_k),
\ee 
where $\tau=t'/t$. Of particular interest are the even and odd scattering wave-functions at $n=1$, $\Gamma_{ek}\equiv\chi_{ek}(1)$ and $\Gamma_{ok}\equiv\chi_{ok}(1)$ given by
\bea\label{eqsmallabkgamma}
\Gamma_{(e/o)k}&=&2e^{i\delta_k^{\pm}}\sin(k+\delta_k^{\pm})\br
&=&\pm {2\over \tau}e^{i\delta_k^{\pm}}\sin \delta_k^{\pm}
=\frac{2\sin k}{{1\mp\tau e^{ik}}}.
\eea
We can use these wave-functions to define a new set of annihilation operators $q_{ek}$ and $q_{ok}$ in terms of which the position-space operators are given by
\be\label{eqqdef}
c_{n}=\frac{1}{\sqrt{2}}\intopi{\frac{dk}{2\pi}\Big(\chi\dn_{ek}(n)q_{ek}\dn+\chi\dn_{ok}(n)q_{ok}\dn\Big)},
\ee
and the tunneling Hamiltonian becomes
\be
H_T=-\frac{1}{\sqrt{2}}\intopi{\frac{dk}{2\pi}[(t_{de}^*\Gamma^{*}_{ek}q\dg_{ek}-t_{do}^*\Gamma^{*}_{ok}q\dg_{ok})d+h.c.]},
\ee
in which the dot is coupled to even/odd scattering states by the 
flux-dependent amplitudes $t_{de}$ and $t_{do}$
\be
\matn{
t_{de}=t_Le^{-i\varphi/2}+t_Re^{i\varphi/2}\vsp\\
t_{do}=t_Le^{-i\varphi/2}-t_Re^{i\varphi/2}\vsp.
}
\ee
The operators $q_{ek}$ and $q_{ok}$ are normalized so as to satisfy the anti-commutation relations
\be
\acom{q\dn_{e/ok_1}}{q\dg_{e/ok_2}}=2\pi\delta(k_1-k_2).
\ee
Another unitary transformation to screening $\psi$ and non-screening $\phi$ channels:
\be
\Psi_k\equiv\mat{\psi_k\\\phi_k}=\mathbb{U}_k\mat{q_{ek}\\q_{ok}} \\ 
\ee
where
\be
\mathbb{U}_k=\frac{1}{\sqrt{2}V_k}\mat{t_{de}\Gamma_{ek}&-t_{do}\Gamma_{ok}\\t_{do}^*\Gamma^{*}_{ok}&\hphantom{-}t_{de}^*\Gamma^{*}_{ek}},\label{equk}
\ee
with the normalization factor $V_k>0$ given by
\bea
V_k=2\sin k\sqrt{t_L^2+t_R^2}\sqrt{\frac{1+\tau^2+2\gamma\tau\cos k\cos\varphi}{(1+\tau^2)^2-4\tau^2\cos^2k}}\qquad\label{eqvk}
\eea
maps the problem to a single channel coupled to an Anderson impurity $H=H_0+H_T+H_d$ with
\bea\label{eqHand}
H_0&=&\intopi{\frac{dk}{2\pi}\eps_k\Big(\psi\dg_k\psi_k\dn+\phi\dg_k\phi_k\dn\Big)}\\
H_T&=&-\intopi{\frac{dk}{2\pi}V_k\Big(\psi\dg_k d+d\dg\psi\dn_k\Big)},
\eea
where the non-screening channel $\phi$ is decoupled from the dot. Here $\eps_k=-2t\cos k$ is the dispersion relation of free electrons and the parameter
\be
\gamma=\frac{2t_Lt_R}{t_L^2+t_R^2}
\ee
characterizes the coupling asymmetry of the dot. Therefore, the problem of the AB ring with an embedded quantum dot is mapped to a single-impurity Anderson model 
with a generally energy-dependent and flux-dependent hybridization parameter, $V_k$. 
\subsection{Kondo model}
An approximation to the Anderson model is given by the Kondo model [\onlinecite{Kondo}]. One can perform the Schrieffer-Wolff transformation [\onlinecite{ShriefferWolf}] to arrive at
\bea\label{eqHK}
H&=&\int{\frac{dk}{2\pi}\eps_k(\psi\dg_k\psi\dn_k+\phi\dg_k\phi\dn_k)}\br
&+&\int{\frac{dkdk'}{(2\pi)^2}(J_{kk'}\psi_k\dg\frac{\vec\sigma}{2}\psi\dn_{k'}\cdot\vec{S}_d+K_{kk'}\psi\dg_k\psi\dn_{k'}}).\qquad
\eea
Here $\vec{S}_d$ is the impurity spin and the couplings $J_{kk'}$ and $K_{kk'}$ are given by [\onlinecite{ShriefferWolf}]
\be
J_{kk'}=V_k(j_k+j_{k'})V_{k'}, \quad
K_{kk'}=V_k(\kappa_k+\kappa_{k'})V_{k'}/2,
\ee
where we have defined 
\be
(j,\kappa)_k=\frac{1}{\eps_k-\eps_d}\pm\frac{1}{U+\eps_d-\eps_{k}}.
\ee
The Kondo model is usually defined with a reduced band width $D\ll t$ so the momentum dependence of 
the couplings can be ignored for the small ABK ring considered in this paper. 
Transport properties are usually given in terms of the diagonal coupling $J_{kk}\propto V_k^2$ which contains 
the first  harmonic of the dimensionless flux, $\phi$. 

\section{Conductance from Kubo formula}
\label{sec:cond}
The current operator may be written as 
\be I=-{e\over 2}{d\over dt}\Delta N ,\ee
where 
\be \Delta N\equiv N_R-N_L\ee
and
\be N_{R/L}=\sum_{n=\pm 1}^{\pm\infty}c\dg_nc\dn_n.\ee
By adding a perturbation to the Hamiltonian, $eV(t)\Delta N/2$, with $V(t)=V_0\cos \Omega t$, the Kubo formula gives the DC conductance. 
\be
G=\frac{e^2}{h}\lim_{\Omega\to 0} {2\pi i \Omega} G'(\Omega )
\label{eqGNN}\ee
where $G'(\Omega )$ is the retarded Green's function of $\Delta N/2$:
\be G'(\Omega )\equiv -{i\over 4}\int_0^\infty dt e^{i(\Omega+i\eta)t}\braket{\com{\Delta N(t)}{\Delta  N(0)}}.\label{G'}
\ee
($\eta$ is an infinitesimal positive convergence factor.)  Alternatively, the same formula for the conductance 
may be obtained by applying a vector potential between the quantum dot and sites $\pm 1$ and 
between sites $+1$ and $-1$.  In some cases it will be convenient to obtain $G'(\Omega )$ by analytic continuation from 
the imaginary time, time-ordered Green's function:
\be \mathcal{G}'(i\omega_p)\equiv -{1\over 4}\int_0^\beta d\tau e^{i\omega_p\tau}\braket{T_\tau \Delta N(\tau )\Delta N(0)}
\ee
where $\beta \equiv \hbar /T$ and $\omega_p\equiv 2\pi p/\beta$. 

\subsection{Kubo formula in terms of Green's functions of screening and non-screening channels}
\noindent
$\Delta N$ can be written in the scattering basis as
\be\label{eqn}
\Delta N=\intopi{\frac{dk_1dk_2}{(2\pi)^2}}
\mat{q_{ek_1}\dg & q_{ok_1}\dg}
\mathbb{A}_{k_1k_2}
\mat{q_{ek_2}\dn \\ q_{ok_2}\dn},
\ee
where the matrices $\mathbb{A}_{k_1k_2}$ are given by
\be\label{eqadef}
\mathbb{A}_{k_1k_2}={1\over 2}\Bigg[\sum_{n=1}^{\infty}-\sum_{-\infty}^{-1}\Bigg]\mat{\chi_{ek_1}^*(n) \\ \chi_{ok_1}^*(n)}\mat{\chi_{ek_2}(n)&\chi_{ok_2}(n)}.
\ee
The off-diagonal matrix elements of $\mathbb{A}_{k_1k_2}$ are partial overlaps of even and odd scattering wave-functions in the left/right leads. A direct summation of \pref{eqadef} using \pref{eqasymchi} and \pref{eqsmallabkphase}
and after introducing appropriate convergence factors yields
\be\label{eqdefa}
\mathbb{A}_{k_1k_2}=2\pi\delta(k_1-k_2)\tau_x+g^R_{k_1}(\eps_{k_2})\mathbb{F}_{k_1k_2},
\ee
where
\be
g^{R/A}_k(\eps)\equiv g_k\dn(\eps\pm i\eta)=\frac{1}{\eps-\eps_k\pm i\eta},
\ee
are the free retarded/advanced Green's functions,
\be
\mathbb{F}_{k_1k_2}\equiv\mat{0&f_{k_1k_2}\\-f^*_{k_2k_1}&0},
\ee
and
\be\label{eqak1k2}
f_{k_1k_2}\equiv 2t'\Gamma^*_{ek_1}\Gamma_{ok_2}\dn.
\ee
We refer to the first term in the left equation of \pref{eqdefa} as the contact term and the second term as the overlap term. 
Here and in the following $\tau_x,\tau_y$ and $\tau_z$ are Pauli matrices in the $\psi-\phi$ basis and $\tau_{\psi}\equiv\frac{1}{2}(\mathbb{1}+\tau_z)$ is the projection operator onto the $\psi$ state.  Using $g^R_{k_1}(\eps_{k_2})^*=-g^R_{k_2}(\eps_{k_1})$, it can be seen that
\be \mathbb{A}_{k_1k_2}=\mathbb{A}_{k_2k_1}^\dagger .\label{Aun}\ee
We will be mainly interested in the diagonal elements of $f$ which
can be written as
\be\label{eqfopt}
\pi\nu_{k} f_{kk}=e^{-i(\delta^+_k-\delta^-_k)}\sin(\delta^+_k-\delta^-_k)
\ee
where $\nu_k$ is the density of states per unit length, per spin, per channel,
\be \nu_{k}\equiv \frac{1}{4\pi t\sin k}.
\ee
$\Delta N$ can be expressed in terms of screening and non-screening channels
\be\label{eqDN}
\Delta N(t)=\intopi{\frac{dk_1dk_2}{(2\pi)^2}}\Psi\dg_{k_1}(t)\mathbb{M}_{k_1k_2}\Psi_{k_2}\dn(t),
\ee
in which the matrix  $\mathbb{M}_{k_1k_2}$ is defined as 
\be\label{eqM}
\mathbb{M}_{k_1k_2}\equiv\mathbb{U}_{k_1}\mathbb{A}_{k_1k_2}\mathbb{U}\dg_{k_2}
\ee
where $\mathbb{U}_k$ is defined in Eq.\,\pref{equk}.  It follows immediately from Eq.\,(\ref{Aun}) that:
\be \mathbb{M}_{k_1k_2}=\mathbb{M}_{k_2k_1}^\dagger \ee
a property which we will use below and which implies that $\Delta N$ is Hermitean. 
\subsection{Conductance formula}\label{sec:disG}
\begin{figure}[tp]
\includegraphics[scale=0.48]{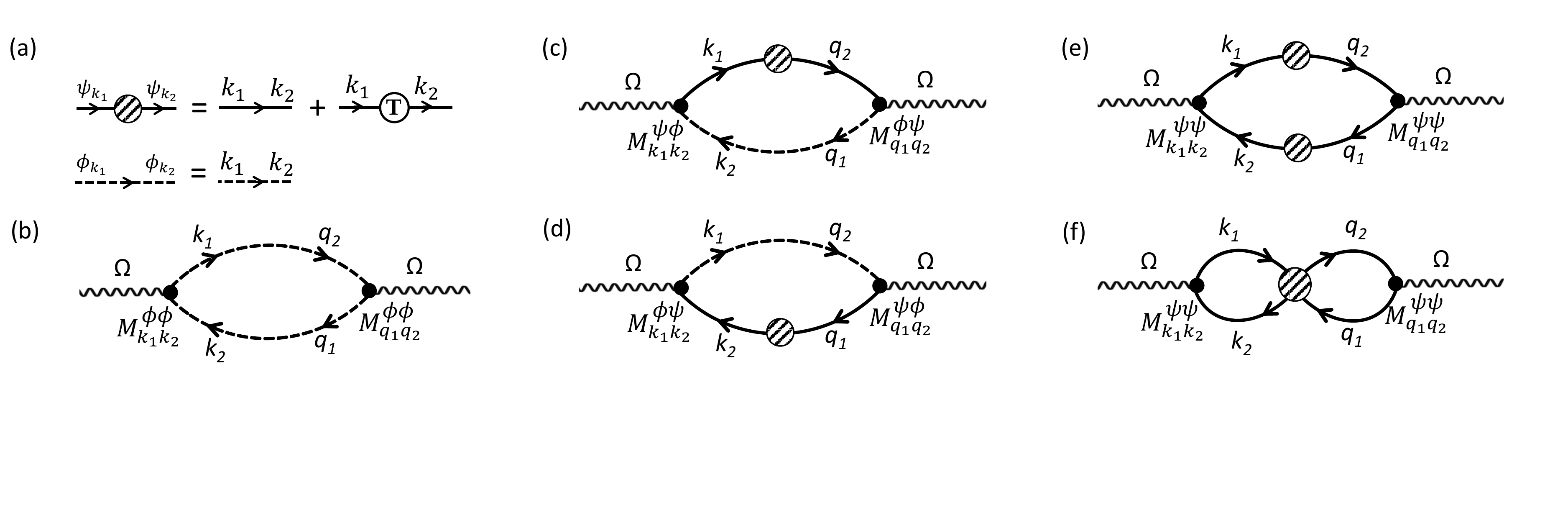}
\caption{\raggedright\small Feynman diagram representation of Kubo calculation of the conductance. Solid and dashed lines represent the free Green's function of the $\psi$ and $\phi$ electrons, respectively, and are diagonal in the initial and final momenta, i.e. proportional to $2\pi\delta(k_1-k_2)$. The hashed circle connecting two solid lines denotes the exact Green's function of $\psi$ electrons. (a) Diagramatic representation of Eq.\,\pref{eqgTg}. A decomposition of the exact $\psi$ Green's function into its free and T-matrix parts is implied in all the other diagrams, as well. Therefore, diagrams (b)-(e) each partially contributes to the background conductance. Diagrams (c)-(e) each partially contribute to the linear in T-matrix part of the conductance. Diagram (e) contributes to the quadratic in T-matrix part of the conductance. (f) The connected part of the conductance.}
\label{fig:Feyncond}
\end{figure}

Inserting Eq.\,\pref{eqDN} into  Eq.\,\pref{G'} we obtain an exact expression for the 
Green's function of $\Delta N$, and hence the conductance, in terms of retarded Green's functions of the single channel Anderson or Kondo model 
and the non-interacting Green's function of the non-screening field $\phi_k$.  Obtaining the retarded Green's function of $\Delta N$ 
from the analytic continuation of the time-ordered imaginary time Green's function, the corresponding Feynman diagrams are drawn in Fig.\,\pref{fig:Feyncond}. Both 2-point and connected 4-point Green's functions occur. 

\subsubsection{Disconnected part}
\label{sec:disc}
The disconnected part of $G'$ is most conveniently dealt with in real time domain where it is  denoted by $G^{'D}$. Using Wick's theorem 
it  can be written:
\bw

\bea\label{eqwick}
G^{'D} (\Omega )=-\frac{i}{2}\intopi{\frac{dk_1dk_2dq_1dq_2}{(2\pi)^4}}\intoinf{dte^{i(\Omega +i\eta )t}}
\tr{
\mathbb{M}\dn_{k_1k_2}\mathbb{G}^>_{k_2q_1}(t)\mathbb{M}\dn_{q_1q_2}\mathbb{G}^<_{q_2k_1}(-t)-
\mathbb{M}\dn_{q_1q_2}\mathbb{G}^>_{q_2k_1}(-t)\mathbb{M}\dn_{k_1k_2}\mathbb{G}^<_{k_2q_1}(t)}\qquad\quad
\eea
\ew

where a factor of 2 from summation over spin indices is  taken into account. Here we used the fact that 
\be \mathbb{G}^<_{k_2q_1\alpha \beta}(t)=\delta_{\alpha \beta}\mathbb{G}^<_{k_2q_1}(t)
\ee
due the SU(2) symmetry of the model. 
Here the  Green's functions $\mathbb{G}_{kq}^>(t)$ and $\mathbb{G}_{qk}^<(t)$ are diagonal matrices in the $\psi-\phi$ space, 
\be
\mathbb{G}^>_{kq}(t)\equiv -i\mat{\brket{\psi\dn_{k}(t)\psi\dg_{q}(0)} & 0 \\ 0 & \brket{\phi\dn_{k}(t)\phi\dg_{q}(0)}},\\
\ee
and
\be
\mathbb{G}^<_{qk}(t)\equiv +i\mat{\brket{\psi\dg_{k}(0)\psi\dn_{q}(t)} & 0 \\ 0 & \brket{\phi\dg_{k}(0)\phi\dn_{q}(t)}}.
\ee
We can write these Green's functions in the Fourier domain,
do the time integral and use the general equilibrium identities $\mathbb{G}^<_{qk}(\omega)=-2if(\omega)\hbox{Im}\mathbb{G}^R_{qk}(\omega)$ and  $\mathbb{G}^>_{kq}(\omega)=2i(1-f(\omega))\hbox{Im}\mathbb{G}^R_{kq}(\omega)$, where $\mathbb{G}^R_{qk}(\omega)$ is 
the matrix retarded single electron Green's function and $f(\omega)=(1+e^{-\beta(\omega-\mu_F)})^{-1}$ is the Fermi distribution.
[Here we used the fact that $\mathbb{G}^R_{kq}(\omega)=\mathbb{G}^R_{qk}(\omega)$ for the Anderson model, as can be seen from Eqs.\,(\ref{eqgTg}) and (\ref{andersontmatrix}).] 
We thus obtain
\bw

\bea
G^{'D}(\Omega )&=&-2i\int \frac{d\omega d\omega'}{(2\pi)^2}[f(\omega)-f(\omega')]\intopi{\frac{dk_1dk_2dq_1dq_2}{(2\pi)^4}}\tr{
\mathbb{M}\dn_{k_1k_2}\hbox{Im}\mathbb{G}^R_{k_2q_1}(\omega')\mathbb{M}\dn_{q_1q_2}\hbox{Im}\mathbb{G}^R_{q_2k_1}(\omega)
}\int_0^\infty dt e^{i(\omega -\omega '+\Omega +i\eta )t}\nonumber \\
&=&2\int \frac{d\omega d\omega'}{(2\pi)^2}{f(\omega)-f(\omega')\over \omega -\omega '+\Omega +i\eta }\intopi{\frac{dk_1dk_2dq_1dq_2}{(2\pi)^4}}\tr{
\mathbb{M}\dn_{k_1k_2}\hbox{Im}\mathbb{G}^R_{k_2q_1}(\omega')\mathbb{M}\dn_{q_1q_2}\hbox{Im}\mathbb{G}^R_{q_2k_1}(\omega)}
\eea
The momentum integral can be seen to be real using $\mathbb{M}_{k_1k_2}=\mathbb{M}_{k_2k_1}^\dagger$ and 
$\hbox{Im}\mathbb{G}^R_{kq}(\omega )=\hbox{Im}\mathbb{G}^R_{qk}(\omega )$. 
Here we focus on the real part of the conductance; 
only the imaginary part of $G'$ contributes to it. 
Thus, we only need
\be \hbox{Im}G^{'D}(\Omega )=-\frac{1}{2\pi} \int_{-\infty}^\infty d\omega [f(\omega)-f(\omega+\Omega)]
\intopi{\frac{dk_1dk_2dq_1dq_2}{(2\pi)^4}}\tr{
\mathbb{M}\dn_{k_1k_2}\hbox{Im}\mathbb{G}^R_{k_2q_1}(\omega +\Omega )\mathbb{M}\dn_{q_1q_2}\hbox{Im}\mathbb{G}^R_{q_2k_1}(\omega)}.
\label{ImG'D}
\ee
Inserting Eq.\,(\ref{ImG'D}) into Eq.\,(\ref{eqGNN}), gives the disconnected part of the DC conductance
\be G^D={2e^2\over h}\int_{-\infty}^\infty d\omega [-f'(\omega )]\mathcal{T}^D(\omega )\label{GDT}\ee
where the disconnected part of the ``transmission probability'' is defined as
\be \mathcal{T}^D(\omega )\equiv \lim_{\Omega \to 0}{\Omega^2\over 2}\intopi{\frac{dk_1dk_2dq_1dq_2}{(2\pi)^4}}\tr{
\mathbb{M}\dn_{k_1k_2}\hbox{Im}\mathbb{G}^R_{k_2q_1}(\omega +\Omega )\mathbb{M}\dn_{q_1q_2}\hbox{Im}\mathbb{G}^R_{q_2k_1}(\omega)}.
\label{trpr}\ee

\ew

The retarded Green's function can be written in terms of the ${\rm T}$-matrix of the Anderson or Kondo model, ${\rm T}_{k_2q_1}(\omega)$:
\be\label{eqgTg}
\mathbb{G}^R_{k_2q_1}(\omega)=2\pi\delta(k_2-q_1)g^{R}_{k_2}(\omega)\mathbb{1}+\tau_{\psi}g^{R}_{k_2}(\omega){\rm T}_{k_2q_1}(\omega)g^{R}_{q_1}(\omega).
\ee
(Note that the $\tau_\psi$ projection matrix implies that the second term is only present for the screening channel, $\psi$.  Also 
note that $\mathbb{G}^R_{k_2q_1\alpha \beta}$, $g^{R}_{k_2,\alpha \beta}$ and ${\rm T}_{k_2q_1\alpha \beta}$ are all $\propto \delta_{\alpha \beta}$ 
due to the SU(2) symmetry of the model. We are suppressing all spin indices.)
For the Anderson model of Eq.\,\pref{eqHand} this T-matrix is related to the retarded Green's function of the dot $G_{dd}^R(\omega)$ via
\be\label{andersontmatrix}
{\rm T}_{k_2q_1}(\omega)={V}_{k_2}G_{dd}^R(\omega){V}_{q_1}.
\ee
We see that $G^D$ is a sum of terms of zeroth, first and second order in the T-matrix.  The  ${\rm T}$-matrix is a smooth function of frequency 
and the needed divergences as $\Omega \to 0$ of the momentum integral in Eq.\,(\ref{trpr}) arise from the singular behaviour of $\mathbb{A}\dn_{q_1q_2}$ 
in Eq.\,(\ref{eqdefa}), $\propto \mathbb{M}\dn_{q_1q_2}$ by Eq.\,(\ref{eqM}), 
and from the factors of $g^R$ in Eq.\,(\ref{eqgTg}).  We find that, after taking the limit $\Omega \to 0$, $\mathcal{T}(\omega)$ is only non-zero 
for $\omega$ inside the band, $|\omega |<2t$. Therefore, it is convenient to write the integration variable $\omega$ in Eq.\,(\ref{GDT}) as $\epsilon_p$:
\be G^D={2e^2\over h}\int_{-2t}^{2t}d\epsilon_p [-f'(\epsilon_p )]\mathcal{T}^D(\epsilon_p ).\label{GDep}\ee
We show in Appendix (\ref{sec:ap02}) that the transmission probability for the disconnected part of the conductance may be written 
in terms of the diagonal on-shell T-matrix of the  Anderson or Kondo model, and the density of states, as
\bea\label{eqTDanderson}
\mathcal{T}^D(\eps_p)&=&\mathcal{T}_0(\eps_p)\vsp\br
&+&\mathcal{Z}_R(\eps_p)\re{ -\pi\nu_p{\rm T}_{pp}(\eps_p)}\vsp\br
&+&\mathcal{Z}_I(\eps_p)\im{-\pi\nu_p{\rm T}_{pp}(\eps_p)}\vsp\br
&+&\mathcal{Z}_2(\eps_p)|-\pi\nu_p{\rm T}_{pp}(\eps_p)|^2\vsp\qqquad
\eea
where
\bea 
\mathcal{T}_0(\eps_p)&=&\frac{4\tau^2\sin^2p}{(1+\tau^2)^2-4\tau^2\cos^2p}\vsp,\label{eqz02}\\
\mathcal{Z}_R(\eps_p)&=&\frac{4\tau\cos p+2\gamma(1+\tau^2)\cos\varphi}{1+\tau^2+2\gamma\tau\cos p\cos\varphi}\sqrt{\mathcal{T}_0[1-\mathcal{T}_0]}
,\qquad\quad\label{eqzr2}\\
\mathcal{Z}_I(\eps_p)&=&1-2\mathcal{T}_0(\eps_p),\label{eqzi}\vsp\\
{\mathcal{Z}}_2(\eps_p)&=&\frac{-(1-\tau^2)^2(1-\gamma^2)+4\gamma^2\tau^2\sin^2p\sin^2\varphi}{[1+\tau^2+2\gamma\tau\cos p\cos\varphi ]^2}
.\label{eqzc2}
\eea
A non-trivial check of Eqs.\,(\ref{GDep})-(\ref{eqzc2}) is the non-interacting ABK ring, $U=0$.  In this 
case the connected 4-point Green's function vanishes, so $\mathcal{T}^D(\eps_p)$ gives the entire transmission probability. In 
App.\,\pref{sec:non-int} we derive Eq.\,(\ref{GDep}) from the Landauer formalism and confirm that Eqs.\,(\ref{eqTDanderson})-(\ref{eqzc2}) give 
the correct transmission probability.

\subsubsection{Connected part}
\label{subcon}
 The connected contribution to ${\mathcal G}'(i\omega_p)$ is given by
\be\label{eqg4p}
\mathcal{G}^{'C}(i\omega_p)={1\over 4}\int_0^\pi \frac{dk_1dk_2dq_1q_2}{(2\pi)^4} M^{11}_{k_1k_2}M^{11}_{q_1q_2}
\mathcal{G}^C_{k_1k_2q_1q_2}(i\omega_p)
\ee
which is a functional of the connected four-point Green's function $\mathcal{G}^C_{k_1k_2q_1q_2}(i\omega_p)$ defined as
\bw

\bea\label{eqp4c}
\mathcal{G}^{C}_{k_1k_2q_1q_2}(i\omega_p)=-\int_0^\beta e^{i\omega_p\tau}d\tau
\sum_{\sigma\sigma'}\braket{
T_\tau\psi\dg_{k_1\sigma}(\tau )\psi\dn_{k_2\sigma}(\tau )\psi\dg_{q_1\sigma'}(0)\psi\dn_{q_2\sigma'}(0)}_{C}.
\eea
The subscript and the superscript $C$ both refer to the connected part. 
 Using equation-of-motion techniques and as is clear from Fig.\,\pref{fig:Fig4}, the connected part of the four-point function can be written in 
 imaginary time domain as
\be\label{eq79}
\mathcal{G}^{C}_{k_1k_2q_1q_2}(\tau)=\intob{d\tau_1d\tau_2d\tau_3d\tau_4 g_{k_2}(\tau-\tau_2)g_{q_2}(0-\tau_4)\mathcal{G}^{amp}_{k_1k_2q_1q_2}(\tau_1,\tau_2,\tau_3,\tau_4)g_{k_1}(\tau_1-\tau)g_{q_1}(\tau_3-0)}
\ee
in terms of the amputated function which is proportional to the connected four-point function of the $d$ electrons
\be\label{eq47}
\mathcal{G}^{amp}_{k_1k_2q_1q_2}(\tau_1,\tau_2,\tau_3,\tau_4)=V_{k_1}V_{k_2}V_{q_1}V_{q_2}\mathcal{G}^C_{dd}(\tau_1,\tau_2,\tau_3,\tau_4)
\ee
where
\be
\mathcal{G}_{dd}^C(\tau_1,\tau_2,\tau_3,\tau_4)\equiv\sum_{\sigma\sigma'}-\braket{T_{\tau}d\dg_{\sigma}(\tau_1)d\dn_{\sigma}(\tau_2)d\dg_{\sigma'}(\tau_3)d\dn_{\sigma'}(\tau_4)}_C.
\ee
Here 
\be g_k(\tau )\equiv [f(\tau )-\theta (\tau )]e^{-\epsilon_k|\tau |}.\ee
%
In Fourier-domain 

\bea\label{eq80}
\mathcal{G}^{C}_{k_1k_2q_1q_2}(i\omega_p)=
\frac{V_{k_1}V_{k_2}V_{q_1}V_{q_2}}{\beta^2}\sum_{mn}g_{k_2}(i\varpi_m+i\omega_p)g_{q_2}(i\varpi_n-i\omega_p)\mathcal{G}^C_{dd}(i\varpi_m,i\varpi_m+i\omega_p,i\varpi_n,i\varpi_n-i\omega_p)g_{k_1}(i\varpi_m)g_{q_1}(i\varpi_n).\nonumber
\eea
\ew

Here $\varpi_m$ and $\varpi_n$ are fermionic and $\omega_p$ is a bosonic Matsubara frequency. Analytic continuation 
to real frequencies, $i\omega_p\ra\Omega+i\eta$ gives the connected part of the retarded four-point function to be plugged into Eq.\,\pref{eqg4p}. 

We now argue that the connected part of the conductance can also be written:
\be G^C={2e^2\over h}\int_{-\infty}^\infty d\omega [-f'(\omega )]\mathcal{T}^C(\omega )\label{GCT}\ee
 in terms of a connected part of the ``transmission probability'' $\mathcal{T}^C(\omega )$. This is an important result 
since it implies that the total conductance at temperatures small compared to the band width is determined by 
universal low energy properties of the system. It is also crucial for approximately eliminating the connected 
term from the conductance at low temperatures, as we show in Sec.\,(\ref{sec:con}). To establish this result it is convenient 
to write $\mathcal{G}^{'C}(i\omega_p)$ in terms of a partially amputated Green's function, $P(i\varpi_m,i\varpi_m+i\omega_p)$:
\be \mathcal{G}^{'C}(i\omega_p)=\frac{1}{\beta}\sum_m P(i\varpi_m,i\varpi_m+i\omega_p)\label{summ}
\ee
where
\bw

\bea 
P(i\varpi_m,i\varpi_m+i\omega_p)\equiv \frac{1}{4}\intopi{\frac{dk_1dk_2}{(2\pi)^2}} 
M^{11}_{k_1k_2} g_{k_1}(i\varpi_m)g_{k_2}(i\varpi_m+i\omega_p)V_{k_1}V_{k_2}\times  \br
&&\hspace{-4cm}
\intob{d\tau_1d\tau_2e^{i\varpi_m\tau_1+i(\varpi_m+\omega_p)\tau_2}\braket{T_{\tau}
d\dg(\tau_1)d\dn(\tau_2)\Delta N(0)}}_C.\label{Pdef}
\eea
\ew

\begin{figure}[htp]
\centering
\includegraphics[scale=.39]{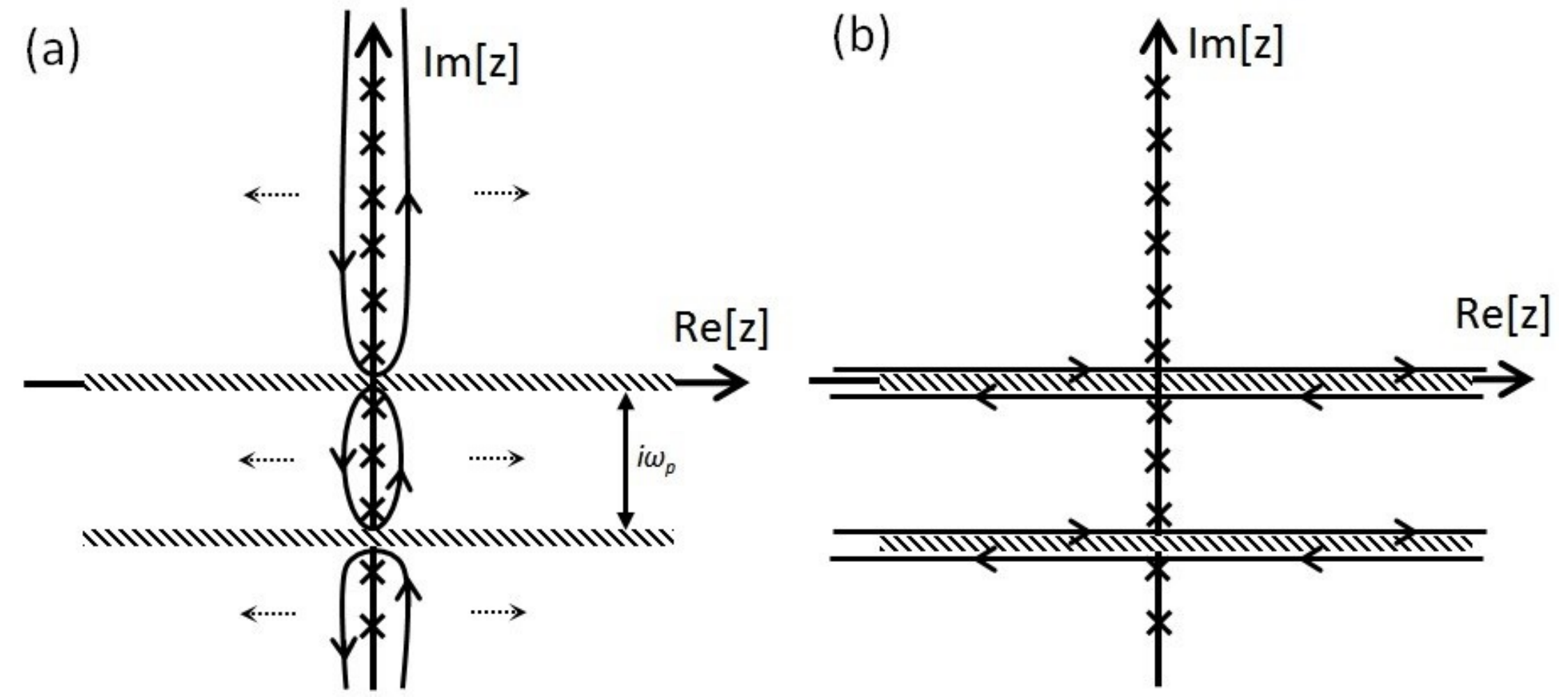}
\caption{\raggedright\small The required contour deformation for summing over Matsubara frequencies in Eq.\,(\ref{summ}). 
The hatched regions mark the vertical position of singularities of $P(z,z+i\omega_p)$.}\label{fig:contour}
\end{figure}

We now consider transforming the sum over $\varpi_m$ in Eq.\,(\ref{summ}) into a contour integral in the complex $z$-plane. 
Thus we must consider the singularities of $P(z,z+i\omega_p)$ for arbitrary complex $z$. We expect these singularities 
to lie along the real $z$-axis and along the line $z=-i\omega_p+x$ for real $x$. 
[See the spectral representation of this function  in (\ref{eqLehQ}) and the discussion thereafter.]
The Fermi distribution has 
poles of residue $-1/\beta$ at $z=i\varpi_m$. Thus we may write the sum in Eq.\,(\ref{summ}) as integrals around the 3 contours 
shown in Fig.\,(\ref{fig:contour}a).  We may then deform these contours into 4 horizontal lines displaced infinitesimally 
above and below the lines $z=x$ and $z=x-i\omega_p$ for $-\infty <x<\infty$ as shown in Fig.\,(\ref{fig:contour}b) giving:
\bea
\mathcal{G}'(i\omega_p)=\intinf{\frac{d\omega}{2\pi i}}f(\omega )&\Big\{&P(\omega+i\eta',\omega+i\omega_p )\br 
&-&\vsp P(\omega-i\eta',\omega+i\omega_p )\br
&+&\vsp P(\omega-i\omega_p ,\omega+i\eta')\br &-&P(\omega-i\omega_p ,\omega-i\eta')
\Big\}.\qquad\label{intP}
\eea

\noindent
(Here $\eta '$ is a positive infinitesimal corresponding to the displacements of the integration lines.) 
Next, we consider the analytic continuation of $\mathcal{G}'(i\omega_p)$ to real frequency:
\be i\omega_p\to \Omega +i\eta ,\qquad \mathcal{G}'\to G',\ee\\

\noindent
where $\eta$ is another positive infinitesimal,with $\eta '\ll \eta \ll 1$. 
Finally, from Eq.\,(\ref{eqGNN}), we must multiply $G'(\Omega )$ by a 
factor of $\Omega$ and take $\Omega \to 0$. It can be seen that the integrals of the first and last terms in Eq.\,(\ref{intP}) 
remain finite in this $\Omega \to 0$ limit. This follows because all singularities of $P(\omega +i\eta ',\omega +\Omega +i\eta)$ 
are below the real $\omega$ axis, at $\omega =E-i\eta '$ or $E-\Omega -i\eta$ where $E$ is the 
difference of energies of two states of the system. This follows from the spectral decomposition:
\bw

\bea\label{eqLehQ}&&
\intob{d\tau_1d\tau_2e^{z_1\tau_1+z_2\tau_2}\braket{T_{\tau}
d(\tau_1)d\dn(\tau_2)\Delta N(0)}}
\br
&&\qqquad=\frac{1}{Z}\sum_{mnp}\bra{p}\Delta N\ket{m}
\Big[
\frac{\bra{m}d\dg\ket{n}\bra{n}d\dn\ket{p}}{z_2+E_n-E_p}\Big(\frac{e^{-\beta E_p}-e^{-\beta E_m}}{z_2-z_1+E_m-E_p}-\frac{e^{-\beta E_n}+e^{-\beta E_m}}{z_1+E_n-E_m}\Big) \nonumber \\ 
&&\qqquad
+\frac{\bra{m}d\dn\ket{n}\bra{n}d\dg\ket{p}}{z_2+E_m-E_n}\Big(\frac{e^{-\beta E_p}+e^{-\beta E_n}}{z_1-E_n+E_p}+\frac{e^{-\beta E_p}-e^{-\beta E_m}}{z_2-z_1+E_m-E_p}\Big)
\Big].\qquad
\eea
\ew

($Z$ is the partition function.) We see that, for $z_1=\omega +i\eta '$, $z_2=\omega +\Omega +i\eta$, all singularities occur below the real $\omega$ axis. 
[Actually we must subtract the disconnected part from Eq.\,(\ref{eqLehQ}) to get a representation of $P(z_1,z_2)$, but this also obeys the desired 
property as mentioned  at the beginning of App. (\ref{quadterm}) using results from App. (\ref{sec:ppi}).]
It thus follows that we may deform the line integral in the $\omega$ plane a finite distance above the real axis 
so that the integral remains finite as $\Omega \to 0$. The same argument applies to 
$P(\omega -\Omega -i\eta ,\omega -i\eta ')$. On the other hand, the integrals of the second and third terms 
in Eq.\,(\ref{intP}) diverge as $\Omega \to 0$, and thus contribute to the conductance. This follows because 
$P(\omega -i\eta ',\omega +\Omega +i\eta )$ has singularities both above and below the real $\omega$-axis 
which can pinch the integration contour as $\Omega \to 0$. Finally, shifting the integration variable $\omega \to \omega +\Omega$, 
in the third term in  Eq.\,(\ref{intP}), we may make the approximation, valid for $\Omega 
\to 0$:
\be G'(\Omega )\approx \int_{-\infty}^\infty {d\omega \over 2\pi i}[f(\omega +\Omega )-f(\omega )]P(\omega -i\eta_1,\omega +\Omega 
+i\eta_2)\label{G'df}\ee
for positive infinitesimals $\eta_1$ and $\eta_2$. For small $\Omega$ we may use 
\be f(\omega +\Omega )-f(\omega )\approx \Omega f'(\omega ).\ee
From Eqs.\,(\ref{eqGNN}) and (\ref{G'df}) we then obtain the connected part of the transmission probability:
\be \mathcal{T}^C(\omega )=\lim_{\Omega \to 0}{\Omega^2\over 8}P(\omega -i\eta_1,\omega +\Omega 
+i\eta_2)+c.c. \ee
where $P$ is defined in Eq.\,(\ref{Pdef}), after analytic continuation to real frequency. 
We again expect that $\mathcal{T}^C(\omega )$ will only be non-zero at $\Omega \to 0$ for $\omega$ inside the energy band, $|\omega |<2t$, so we may again replace the integration variable $\omega$ by $\epsilon_p$. 
Note that our analysis of the connected part of the transmission probability is less complete  that of the 
disconnected part where we were able to explicitly take the limit $\Omega \to 0$ and express $\mathcal{T}^D$ 
in terms of smooth functions. For the connected part, we have not so far been able to accomplish this. See however 
subsection (\ref{useTP}).
We confirm that the connected 
part of the conductance can be written in terms of a transmission probability in our perturbation calculation 
in Sec.\,\pref{sec:pert}.  

\section{Perturbative calculation}
\label{sec:pert}
In this section, we calculate the conductance perturbatively to order $J^2\propto V_k^4/U^2$ a result which 
should be valid for $T\gg T_K$.
\subsection{Disconnected Part: T-matrix of $\psi$ electrons}
\begin{figure}
\centering
\includegraphics[scale=0.37]{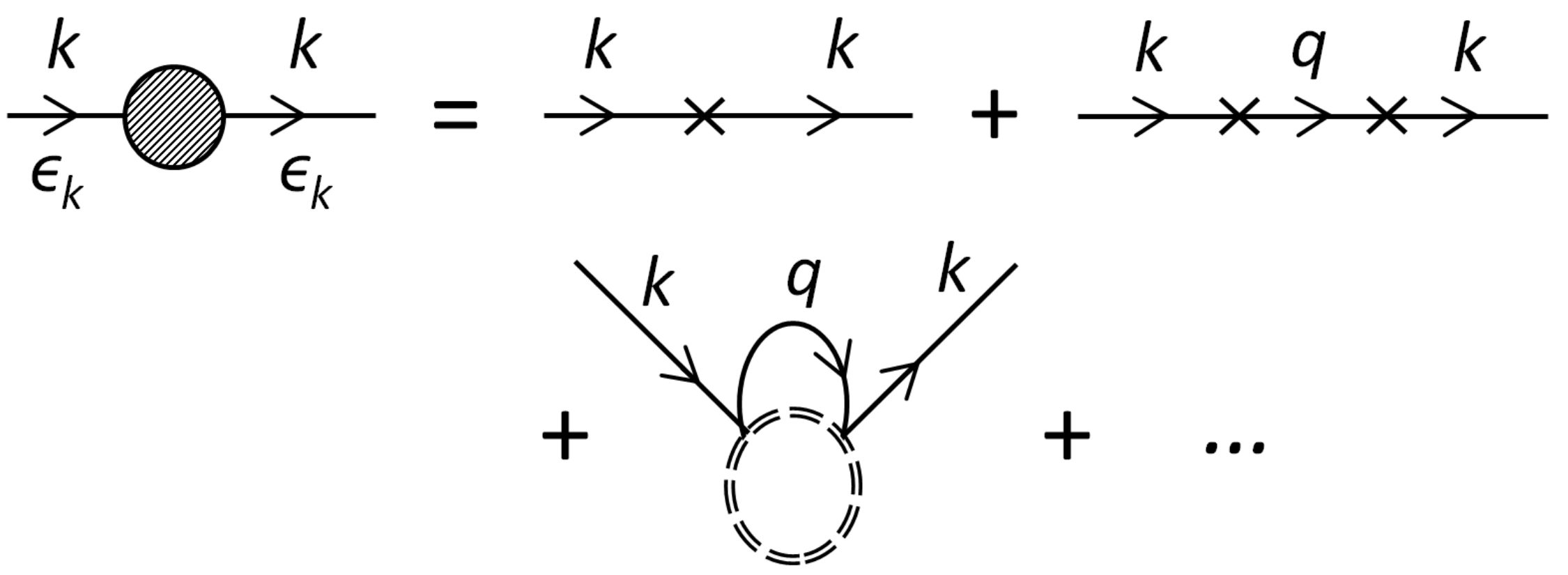}
\caption{\raggedright\small Feynman diagrams for the T-matrix of the Kondo model to order $J^2$. The cross indicates potential scattering and the double dashed line is the `propagator' of the impurity spin. The latter is shown as a loop to emphasize that we trace over the impurity spin. The corresponding hole diagrams are obtained by changing the direction of the arrows.}
\label{fig:fig3}
\end{figure}

\noindent
The relevant Feynman diagrams for the (diagonal element of the retarded on-shell) T-matrix are shown in Fig.\,\pref{fig:fig3} and are given by
\bea\label{eqtkondo}
{\rm T}_{pp}(\eps_p+i\eta)&=&K_{pp}+\int_{\eps_F-D}^{\eps_F+D}{\nu_qd\eps_q\frac{K_{qp}^2}{\eps_p-\eps_q+i\eta}}\qquad\qquad\\
&&\hspace{-1.2cm}+\int_{\eps_F-D}^{\eps_F+D}{\nu_qd\eps_q\frac{\frac{3}{16}J_{qp}^2}{\eps_p-\eps_q+i\eta}}+O(J^3\propto V^6).
\eea
The factor of $3/16$ multiplying the third term comes from the correlation function of the impurity spin. 
At lowest order in $J$ we may use the free spin Green's function:
\be \braket{T_\tau S^a_d(\tau_1)S^b_d(\tau_2)}={1\over 4}\delta^{ab}\ee
independent of $\tau_1$ and $\tau_2$, yielding:
\be \sum_{\beta}\braket{T_{\tau}(\vec{S}_d(\tau_1).\vec{\sigma}_{\alpha\beta})(\vec{S}_d(\tau_2).\vec{\sigma}_{\beta\gamma})}=
{3\over 4}\delta_{\alpha \gamma}.\label{eqspinsum}\ee
The first two terms in Eq.\,\pref{eqtkondo} are potential scattering terms that satisfy the optical theorem 
\be\label{eqt}
\abs{-\pi\nu_p{\rm T}_{pp}(\eps_p)}^2=\im{-\pi\nu_p{\rm T}_{pp}(\eps_p)}.
\ee
 They depend on the position of the dot level $\eps_d$ and can be set to zero ($K_{pp}\approx 0$) by tuning the dot to the middle of two Coulomb resonances $\eps_d-\eps_F\approx-U/2$. The third term contains both real and imaginary parts. The real part depends on the details of the conduction band. This term shows that the conductance is determined by the properties of the system not only at energies close to the Fermi energy but all energies over the full reduced band $\pm D$; it introduces non-universalities that limit the predictive power of the Kondo model. Generally, for energies much smaller than the original band width ($D\ll t$) the total S-matrix can be written as a product from spin and charge sectors [\onlinecite{Affleck93}] which implies that the single-particle T-matrix takes the form [\onlinecite{Affleck08}]
\be
-2\pi\nu{\rm T}_{pp}(\eps_p,T)=-2\pi\nu {\rm T}^K_{pp}(\eps_p,T)e^{2i\delta}+i(e^{2i\delta}-1).
\ee
Here ${\rm T}^K_{pp}(\eps_p)$ corresponds to the T-matrix of a particle-hole symmetric Kondo Hamiltonian 
(for example, the model considered here with $\tau =0$ and $p_F=\pi /2$) 
and is purely imaginary to order $J^2$. $\delta$ corresponds to the total phase shift at the Fermi energy induced by all potential scattering sources that break particle-hole symmetry and is a complicated function of all the parameters of the model. Thus, the $O(J^2)$ term in the real part of the T-matrix is expected to merely contribute to $\delta$ at low temperatures. Although these potential scatterings contribute to the AB oscillations in the conductance, they do not have a strong  dependence 
on energy or temperature and are not relevant for the Kondo physics. Therefore, they will be neglected, $\delta\ra 0$, in the following. To order $J^2$ the rest is
\be\label{eqtmatrix}
-\pi\nu_p{\rm T}_{pp}^K(\eps_p)=i\frac{3\pi^2}{16}\nu_p^2J_{pp}^2.
\ee
Note that, to $O(J^2)$,  ${\rm T}^K_{pp}$ is purely imaginary so that the $\hbox{Re}[{\rm T^K}]$ and $|{\rm T}^K|^2$ terms in Eq.\,(\ref{eqTDanderson}) don't contribute.  \\

\subsection{Connected Part}
\begin{figure}[h!]
\centering
\includegraphics[scale=0.20]{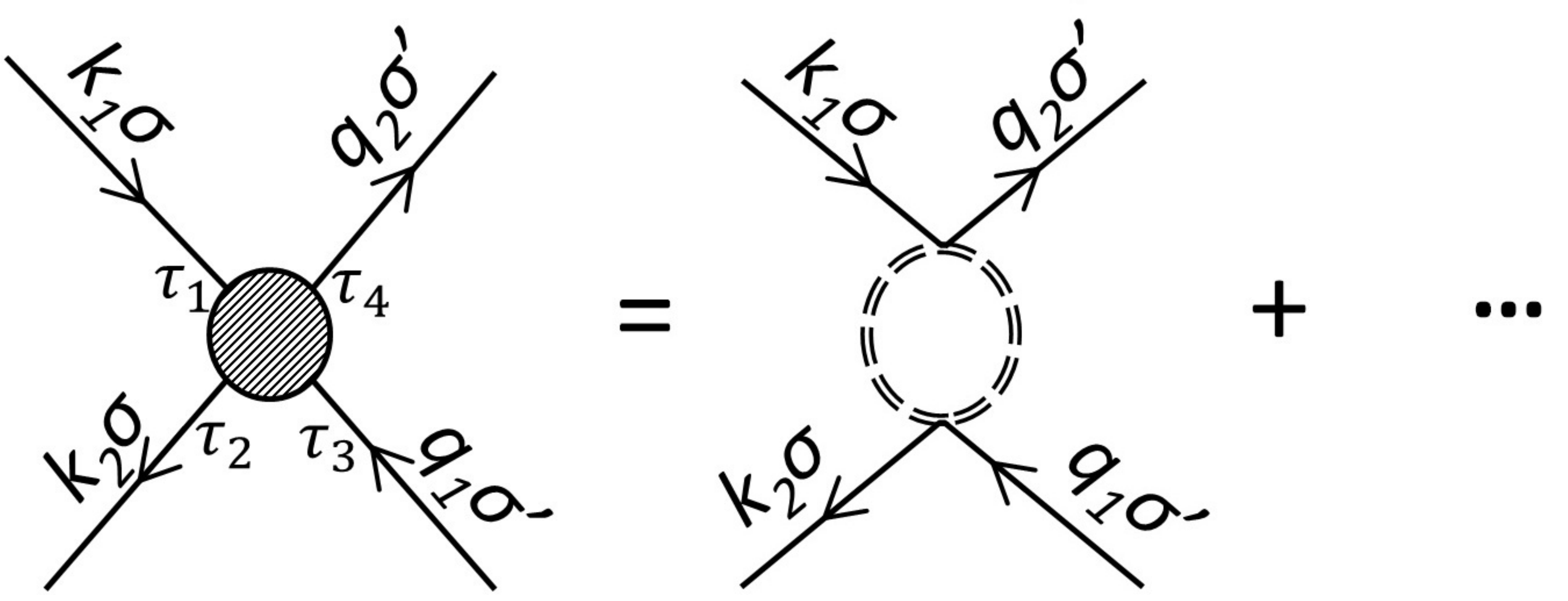}
\caption{\raggedright\small The amputated connected function to order $J^2$. Note that in contrast to some previous works [\onlinecite{Kaminski}], here we trace over the impurity spin.}\label{fig:Fig4}
\end{figure}

\noindent
To order $J^2$ the relevant contribution is given by the Feynman diagram shown in Fig.\,\pref{fig:Fig4}.
\\
\bw
\noindent
The vertex function is given by
\be
\mathcal{G}_{k_1k_2q_1q_2}^{amp}(\tau_1,\tau_2,\tau_3,\tau_4)=\frac{1}{4}J_{q_2k_1}J_{k_2q_1}\sum_{\alpha\beta}\braket{T_{\tau}(\vec{S}_d(\tau_1).\vec{\sigma}_{\alpha\beta})(\vec{S}_d(\tau_2).\vec{\sigma}_{\beta\alpha})}\delta(\tau_1-\tau_4)\delta(\tau_2-\tau_3).
\ee
Using the result of Eq.\,(\ref{eqspinsum}) and 
  Fourier transforming we get
\be\label{eqgcamp0}
\mathcal{G}_{k_1k_2q_1q_2}^{amp}(i\varpi_m,i\varpi_{m}+i\omega_p,i\varpi_n,i\varpi_n-i\omega_p)=\frac{3}{8}J_{q_2k_1}J_{k_2q_1}\delta_{i\varpi_m+i\omega_p-i\varpi_n,0}.
\ee
Plugging this into Eq.\,\pref{eq79} we have
\be
\mathcal{G}^{C}_{k_1k_2q_1q_2}(i\omega_p)=\frac{3}{8}J_{q_2k_1}J_{k_2q_1}\frac{1}{\beta}\sum_m{g_{k_2}(i\varpi_m+i\omega_p)g_{q_2}(i\varpi_m)g_{k_1}(i\varpi_m)g_{q_1}(i\varpi_m+i\omega_p)}.
\ee
The vertex function does not depend on energy and
we can express the summation over Matsubara frequencies as a contour integral and deform the contour as sketched in Fig.\,\pref{fig:contour}.
 After analytic continuation, $i\omega_p\ra\Omega+i\eta$, we write the result as an integral over real frequency
\bea
\mathcal{G}^{C}_{k_1k_2q_1q_2}(\Omega+i\eta)=\frac{3J_{q_2k_1}J_{k_2q_1}}{16\pi i}\intinf{d\omega}\Big\{
&+&f(\omega)\Big[g^R_{k_2}
(\omega+\Omega)g^R_{q_2}(\omega)
g^R_{k_1}(\omega)g^R_{q_1}(\omega+\Omega)\Big]\br
&-&f(\omega)\Big[
g_{k_2}^R(\omega+\Omega)g_{q_2}^A(\omega)
g_{k_1}^A(\omega)g_{q_1}^R(\omega+\Omega)\Big]\br
&+&f(\omega)\Big[
g_{k_2}^R(\omega)g_{q_2}^A(\omega-\Omega)
g_{k_1}^A(\omega-\Omega)g_{q_1}^R(\omega)\Big]\br
&-&f(\omega)\Big[
g_{k_2}^A(\omega)g_{q_2}^A(\omega-\Omega)
g_{k_1}^A(\omega-\Omega)g_{q_1}^A(\omega)\Big]\Big\}.
\eea
Note that this is a special case of the general result discussed in Sub-Section \pref{subcon}. In this simple case the validity of this expression can be checked explicitly. 
Terms that contain all retarded or all advanced propagators do not contribute to the DC conductance. So we drop the first and the last lines and shift 
the integration variable by $\Omega$ in the third line to obtain, after Taylor expanding $f(\omega -\Omega )$, 
\be
\mathcal{G}^{C}_{k_1k_2q_1q_2}(\Omega+i\eta)\ra\frac{3J_{q_2k_1}J_{k_2q_1}}{16\pi i}\intinf{d\omega}\Omega f'(\omega)\Big[g_{k_1}^*(\omega)g_{k_2}\dn(\omega+\Omega)g_{q_1}\dn(\omega+\Omega)g_{q_2}^*(\omega)\Big].
\ee
Inserting this into Eq.\,\pref{eqg4p} we can write the connected part of the conductance as
\bea
G^{C}
&=&\frac{-e^2}{4h}\lim_{\Omega\ra 0}\int{d\eps_p}f'(\eps_p)\intopi{\frac{dk_1dk_2dq_1dq_2}{(2\pi)^4}}\frac{3J_{q_2k_1}J_{k_2q_1}}{8V_{k_1}V_{k_2}V_{q_1}V_{q_2}}{I_{C}(k_1,k_2;\eps_p,\Omega )}{I'_{C}(q_1,q_2;\eps_p,\Omega )}
\eea
where the two functions $I_{C}$ and $I'_{C}$ are defined as
\bea
I_{C}(k_1,k_2;\eps_p,\Omega )=\Omega V_{k_1}V_{k_2}M^{\psi\psi}_{k_1k_2}g_{k_1}^*(\eps_p)g_{k_2}\dn(\eps_p+\Omega),%
\quad
I'_{C}(q_1,q_2;\eps_p,\Omega )=\Omega V_{q_1}V_{q_2}M^{\psi\psi}_{q_1q_2}g_{q_1}\dn(\eps_p+\Omega)g_{q_2}^*(\eps_p).\qquad
\eea

\ew

\begin{figure}
\includegraphics[scale=0.5]{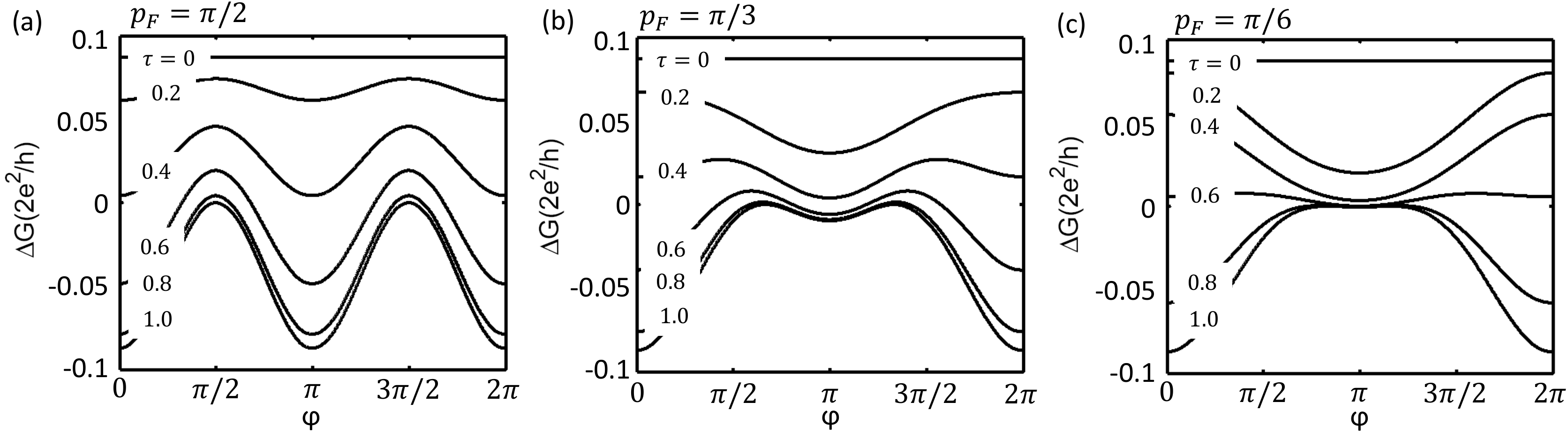}
\caption{\small (a-c) Kondo-type corrections to the conductance versus flux due to the presence of the dot for $T_K\ll T\ll t$ for various values of $p_F$ (proportional to density) and background tunneling amplitude $\tau=t'/t$. 
We have chosen $\gamma =1$ and adjusted $(t_L^2+t_R^2)/(Ut)$ to a different value for each curve, given by (\ref{tLRcond}), so that $\nu_{p_F}J_{p_Fp_F}$ 
has the value $.217$ at $\varphi =0$ for each curve. 
}\label{fig:results}
\end{figure}

Taking the limit $\Omega \to 0$ of such functions is explained in Appendix \pref{sec:ppi}. 
Indeed, these functions are very similar to the matrices $\mathbb{I}_{D2}$ and $\mathbb{I}_{D2}'$ and  the needed propagator-product identities are similar to those of Eqs.\,\pref{eqID2} and \pref{eqID2p} apart from some extra factors of $2(2\pi)^2\delta(p-k_1)\delta(p-k_2)$ and $2(2\pi)^2\delta(p-q_1)\delta(p-q_2)$.
Therefore the connected four-point contribution to the conductance can be written in terms of a transmission probability
\be\label{eqtpc}
G^C=\frac{2e^2}{h}\int{d\eps_p[-f'(\eps_p)]\mathcal{T}^{C}(\eps_p)},
\ee
where
\be
\mathcal{T}^{C}(\eps_p)=\mathcal{Z}_2(\eps_p){3\pi^2\over 16}\nu_p^2J_{pp}^2
\ee         
and $\mathcal{Z}_2(\epsilon_p)$ is given in Eq.\,(\ref{eqzc2}).
Note that the disconnected and connected parts of the transmission probability are the same order of magnitude.

\subsection{total conductance}
\noindent
The total conductance to order $O(J^2)$ is given by combining the connected and disconnected parts:
\be\label{eqfinal0}
G=\int{d\eps_p[-f'(\eps_p)]\mathcal{T}_0(\eps_p)}+\Delta G,
\ee
where the corrections to the conductance due to the presence of the dot $\Delta G$ are equal to
\bea
\Delta G&=&\frac{2e^2}{h}\int{d\eps_p}[-f'(\eps_p)]
\left[\mathcal{Z}_2(\eps_p)+\mathcal{Z}_I(\eps_p)\right]\frac{3\pi^2}{16}(\nu_p J_{pp})^2,\br\label{eqfinal}
\eea
where $\mathcal{Z}_2(\eps_p)$ and $\mathcal{Z}_I(\eps_p)$, proportional to connected and disconnected parts,  are given in Eqs.\,(\ref{eqzc2}) and (\ref{eqzi}) respectively.
The  Kondo coupling factor is equal to
\bea\label{eqnuJ}
&&\frac{3\pi^2}{16}(\nu_pJ_{pp})^2=\br 
&&12\sin^2p\Big(\frac{t_L^2+t_R^2}{Ut}\Big)^2\Big(\frac{1+\tau^2+2\gamma\tau\cos p\cos\varphi}{(1+\tau^2)^2-4\tau^2\cos^2p}\Big)^2.\qquad
\eea
At temperatures small compared to the band width but large compared to $T_K$, considered here, the integration over $f'(\eps_p)$ does not introduce a considerable thermal smearing and we can replace $\eps_p$ with $\eps_F$ in the rest of the integrand, yielding
\be \Delta G=\frac{2e^2}{h}
\left[\mathcal{Z}_2(\eps_F)+\mathcal{Z}_I(\eps_F)\right]\frac{3\pi^2}{16}(\nu_{p_F} J_{p_Fp_F})^2.\ee
 The flux-dependence of the conductance has two origins. One is through the flux-dependence of the Kondo coupling $J$ which affects the flux dependence of the Kondo temperature [\onlinecite{MA}, \onlinecite{Yoshii}]. The other source of flux dependence is through  $\mathcal{Z}_2$.

Of course, Eq.\,\pref{eqfinal} is just the result of perturbation theory to $O(J^2)$. We expect that higher order terms will renormalize the Kondo coupling, giving it a temperature dependence:
\be
J_{p_Fp_F}(T)=J_{p_Fp_F}+\nu_{p_F}J^2_{p_Fp_F}\ln(D/T)+\cdots ,
\ee
where $D$ is of order the band width, $t$. 
The simple form of this renormalized coupling comes from the fact that for the small ring considered here, $J_{kq}$ is a slowly varying function of energy 
on scales of order the band width. 
The effective Kondo coupling thus grows large at the Kondo temperature
\be
T_K\approx De^{-1/(\nu_{p_F} J_{p_Fp_F})}.
\ee
Thus our perturbative result should only be reliable at $T\gg T_K$. In this regime we may write:
\be \nu_{p_F} J_{p_Fp_F}(T)\approx {1\over \ln (T/T_K)}.\ee
 In the opposite limit, $T\ll T_K$ we can use the results of [\onlinecite{MA}, \onlinecite{Yoshii}] based on Fermi liquid theory and NRG.

Fig.\,\pref{fig:results} shows $\Delta G$ versus $\varphi$ for $\gamma =1$ and  various values of  $\tau=t'/t$,  
electron density $2p_F/\pi$  and $(t_L^2+t_R^2)/(Ut)$ at $T_K\ll T\ll t$.
We have adjusted the value of $(t_L^2+t_R^2)/(Ut)$, as we adjust $p_F$ and $\tau$, so that  $\nu_{p_F}J_{p_Fp_F}$ has the maximum value $.217$
 for each curve (occurring for $\varphi =0$),  small 
enough we hope for perturbation theory to be valid.  This corresponds to the condition
\bea \sin p_F\cdot \frac{t_L^2+t_R^2}{Ut}\cdot 
\frac{1+\tau^2+2\tau\cos p_F}{(1+\tau^2)^2-4\tau^2\cos^2p_F}&=&{\pi \over 8}\times .217\br
&\approx& .0852.\qquad\label{tLRcond}
\eea
At $p_F=\pi /2$, the flux-dependence of the Kondo coupling, $J_{k_Fk_F}$ vanishes and the AB oscillations originate from the flux-dependence of the coefficient $\mathcal{Z}_2(\eps_k)$ in Eq.\,\pref{eqzc2} which only contains the second harmonic of $\varphi$ at this density. 
This can be seen in Fig.\,(\ref{fig:results}a), while Figs.\,[\ref{fig:results}(b,c)] show these corrections for lower densities, i.e. $p_F=\pi/3$ and $\pi/6$. We need to emphasize that the corrections $\Delta G$ shown are only the part containing the imaginary part of the T-matrix which is universal. At low densities, the flux-dependence of the Kondo coupling becomes important and higher harmonics create a plateau-like feature in the conductance as a function of $\varphi$. 

Note that for symmetric coupling ($\gamma=1$) and at zero flux ($\varphi=0$), we have $\mathcal{Z}_2=0$ and the sign of Kondo-type conductance correction is set by the sign of $\mathcal{Z}_I=1-2\mathcal{T}_0$. Thus, introducing the dot leads to an enhancement (suppression) of the conductance for $\mathcal{T}_0<0.5$ ($\mathcal{T}_0>0.5$). It can be shown that this is a rather general criteria for parity-symmetric networks with an embedded quantum dot [\onlinecite{KZA}] and it might be related to similar scenarios in transport through molecular junctions with vibrational modes [\onlinecite{Entin09}]. For the present model, the transition happens at $\tau\approx 0.414$ at half-filling ($p_F=\pi/2$) as can be seen in Fig.\,(\ref{fig:results}a).

\section{Eliminating the connected 4 point function from the conductance}
\label{sec:con}
A well-known result of Meir and Wingreen,\cite{Meir} based on Keldysh formalism, shows that for quite general models of interacting quantum dots 
connected to non-interacting leads, the conductance can be expressed in terms of the ${\rm T}$-matrix.  [\onlinecite{Hofstetter}] and [\onlinecite{Yoshii}]
also assumed this.  Thus it is perhaps surprising that our formula for the conductance includes a contribution from a connected 4-point Green's function 
of the Anderson or Kondo model. The Meir-Wingreen argument is based on the fact that the source-drain voltage could be applied with an 
arbitrary asymmetry parameter, $y$, between left and right sides of the quantum dot and the same current should result. 
 In the linear response regime, considered here, Keldysh and Kubo formalisms should yield identical results. 
In   sub-section (\ref{MW}) we recast the Meir-Wingreen argument in Kubo formalism and show that it does straightforwardly allow for exact elimination 
of the connected part in special cases:
 for no reference arm, $t'=0$,  for parity symmetry, $t_L=t_R$, $\varphi =0$ or for $t_L$ or $t_R=0$.  
In sub-section (\ref{useTP}) we use the fact, established in Sec.\,\pref{sec:cond}, that both disconnected and connected parts of the conductance 
can be written as  integrals over energy, $\epsilon_p$, of $f'(\epsilon_p)$ multiplied by ``transmission probabilities'', $\mathcal{T}^D(\epsilon_p)$ 
and $\mathcal{T}^C(\epsilon_p)$.  
Let us denote  the disconnected/connected term 
in the transmission probability, when the Kubo formula is used with $\Delta N/2$ replaced by $N_y$ in Eq.\,(\ref{G'}), by  $\mathcal{T}^{yD/C}(\epsilon_p)$.
We show that if the total  transmission probability $\mathcal{T}^{yC}(\epsilon_p)+\mathcal{T}^{yD}(\epsilon_p)$, is 
assumed to be independent of $y$ at all $\epsilon_p$, then the connected part can be eliminated and the conductance expressed as a 
sum of terms of zeroth and first order in the ${\rm T}$-matrix only.   
We show that this strong assumption holds in lowest order perturbation theory. 
However, it appears unlikely that it holds exactly. 
In sub-section (\ref{approx}) 
we show that the connected part can be approximately eliminated, for temperatures 
small compared to the band width, using only the $y$-independence of the  total conductance, the energy integral of 
$f'(\epsilon_p)[\mathcal{T}^{yC}(\epsilon_p)+\mathcal{T}^{yD}(\epsilon_p)]$. 
  However, while this elimination is possible for the short ABK ring considered here, we argue that 
it would fail for a long ABK ring of length $L$ except at extremely small temperatures, less than the finite size gap of the ring, $\approx v_F/L$.  In the final 
sub-section we apply Keldysh formalism to the ABK ring and again show that the Meir-Wingreen argument does not apply exactly.  We show 
that it  may apply  approximately, for temperatures small compared to the band width (and finite size gap) subject to a plausible assumption 
about a non-equilibrium Green's function.

\subsection{Meir-Wingreen argument recast in Kubo formalism}
\label{MW}
In Sec.\,\pref{sec:cond} we derived the Kubo formula by adding an infinitesimal perturbation $eV(t)\Delta N/2\equiv  eV(t)(N_R-N_L)/2$ to the Hamiltonian. (Here 
\be V(t)=V_0\cos \Omega t\ee
and we eventually take the limit $\Omega \to 0$.)
In the linear response regime, it should be equivalent to apply the voltage asymmetrically, adding  $eV(t)N_y$  to the Hamiltonian, where
\be N_y\equiv yN_R-(1-y)N_L.\label{Nydef}\ee
In the special case, $y=1/2$, $N_y\to \Delta N/2$ reducing to the case considered in Sec.\,\pref{sec:cond}. Note that, in general
\be N_y=\Delta N/2 +(y-1/2)(N_R+N_L).\ee
For the Kondo model, $N_R+N_L$ is the total charge and commutes with the Hamiltonian and with $\Delta N$. Therefore it is easily proven 
that the Kubo formula with $\Delta N/2$ replaced by $N_y$ gives the same conductance as the Kubo formula with $\Delta N/2$. We also 
 expect this to be true for the Anderson model in the parameter range where charge fluctuations of the quantum dot can be ignored at low 
 energies. Applying the source-drain voltage asymmetrically, $eV(t)N_y$, is equivalent to applying an asymmetric vector potential to 
 the links between the quantum dot and sites $\pm 1$ (and between sites $1$ and $-1$). To apply Kubo formalism for 
arbitrary $y$, the simplest approach is to define the current as $I=dN_y/dt$ and measure it in linear response to the 
perturbation $eV(t)N_y$. 
 
 As Meir and Wingreen observed, it may be possible for some models to choose a convenient value of the parameter $y$ so 
 that the connected part is exactly eliminated and the conductance calculation is thus simplified.  Unfortunately, that does not appear to work for the ABK ring
 for general values of the parameters. Let us see why that is so. We now calculate 
 the conductance via the Kubo formula, Eq.\,(\ref{eqGNN}), but with $\Delta N/2$ replaced by $N_y$, for an arbitrary real parameter $y$, 
 in Eq.\,(\ref{G'}).  We next express $N_y$ in the screening, non-screening basis:
\be\label{eqDNy}
N_y(t)=\frac{1}{2}\intopi{\frac{dk_1dk_2}{(2\pi)^2}}\Psi\dg_{k_1}(t)\mathbb{M}^y_{k_1k_2}\Psi_{k_2}\dn(t),
\ee
where $\mathbb{M}^y_{k_1k_2}$ is again expressed in terms of the unitary matrix $\mathbb{U}_k$ (which is independent of $y$)
and a matrix $\mathbb{A}^y_{k_1k_2}$ by
\be\label{eqMy}
\mathbb{M}^y_{k_1k_2}\equiv\mathbb{U}_{k_1}\mathbb{A}^y_{k_1k_2}\mathbb{U}\dg_{k_2}.
\ee
The matrix $\mathbb{A}^y_{k_1k_2}$ is only modified by a shift of the ``contact term'' proportional to the unit matrix:
 \be \mathbb{A}^y_{k_1k_2}=\mathbb{A}^{1/2}_{k_1k_2}+(2y-1)2\pi \delta (k_1-k_2)\mathbb{1}\label{Ay}\ee
 where $\mathbb{A}^{1/2}_{k_1k_2}$ is the quantity simply denoted as $\mathbb{A}_{k_1k_2}$ in Eq.\,(\ref{eqdefa}). 
 The connected part of the conductance is again given by Eq.\,(\ref{eqg4p}) with $\mathbb{M}^{11}_{k_1k_2}$ replaced by
 $\mathbb{M}^{y11}_{k_1k_2}$. Thus the connected part of the conductance will be eliminated if it is possible to choose 
 the real parameter $y$ so that 
 \be \mathbb{M}^{y11}_{k_1k_2}=0\ee
 for all $k_1$ and $k_2$. Unfortunately, since $\mathbb{M}^{y11}_{k_1k_2}$ generally depends non-trivially on $k_1$ and $k_2$, 
 this is usually not possible. 
 
 An exception occurs for the case of no reference arm, $\tau =0$. Then the matrix $\mathbb{U}_{k}$ becomes independent of $k$, the 
 overlap term in $ \mathbb{A}^y_{k_1k_2}$ vanishes and $ \mathbb{M}^{y11}_{k_1k_2}$ simplifies to:
 \be  \mathbb{M}^{y11}_{k_1k_2}=2\pi \delta (k_1-k_2)
 [2y-1-\sqrt{1-\gamma^2}].
 \ee
(Here we assume, without loss of generality, that $t_L>t_R$.) 
 Thus $ \mathbb{M}^{y11}_{k_1k_2}=0$ for all $k_1$ and $k_2$ when we choose:
 \be y={1\over 2}\left(1+\sqrt{1-\gamma^2}\right)={t_L^2\over t_L^2+t_R^2}.\label{yspnra}\ee
The vanishing of $ \mathbb{M}^{y11}_{k_1k_2}$ also implies that the disconnected term quadratic in the $\mathcal{T}$-matrix vanishes. 
We extend the calculation of the $\mathcal{Z}$ coefficients to general $y$ in App.\,(\ref{geny}). 
 Eqs.\,(\ref{eqzr2}), (\ref{eqzi2y}) and (\ref{eqzc2y}) then give 
\be \mathcal{Z}^y_R= \mathcal{Z}^y_2=0,\ \  \mathcal{Z}_I^y=\gamma^2\ee 
for the special value of $y$ in Eq.\,(\ref{yspnra}). Thus the conductance can be written:
\be G=\frac{2e^2}{h}\gamma^2\int{d\eps_p}[-f'(\eps_p)]\im{-\pi\nu_p{\rm T}(\eps_p)}.\label{Gt0}\ee
This result has been already obtained in [\onlinecite{Meir}, \onlinecite{Pustilnik}, \onlinecite{Pustilnik04}, \onlinecite{Carmi}].

Another way of understanding what is so much simpler about the case of no reference arm is to observe that,  even for the symmetric case $y=1/2$, the term 
in $\Delta N$ quadratic in $\psi$ has the simple form:
\be \Delta N_{\psi \psi}=-\sqrt{1-\gamma^2}\int_0^\pi {dk\over 2\pi}\psi^\dagger_k\psi_k\equiv  -\sqrt{1-\gamma^2}N_\psi .\ee
$N_\psi$, the total number of screening electrons,  is an exactly conserved quantity in the Kondo model and approximately conserved at low energies in the Anderson model 
in the regime where charge fluctuations of the dot can be ignored. It then follows that the total contribution to $G'(\Omega )$,
defined in Eq.\,(\ref{G'}) which is quartic in $\psi$ is:
\be G'_{4}(\Omega )\equiv -{i(1-\gamma^2)\over 4}\int_0^\infty dt e^{i(\Omega +i\eta )t}\braket{[N_\psi (t),N_\psi (0)]}=0\label{0}\ee
since $N_\psi (t)=N_\psi (0)$ and hence $[N_\psi(t),N_\psi (0)]= [N_\psi(0),N_\psi (0)]=0$. [These equations, and Eq.\,(\ref{GCyt0}) below, are exact equalities  for the Kondo model and  approximate ones for the Anderson model.]
Interestingly, the connected part of the Green's function in Eq.\,(\ref{0}) is non-zero.  Rather, the conservation of $N_\psi$ implies 
  a ``Ward identity'' [\onlinecite{Ward}] relating the connected part to the disconnected part of linear and quadratic order in the ${\rm T}$-matrix.  
  By comparing the disconnected part at $y=1/2$, and $\tau =0$, determined by Eqs.\,(\ref{eqTDanderson}) to (\ref{eqzc2}),
 with the exact conductance given in Eq.\,(\ref{Gt0}), 
  we see that the connected part of the conductance, for the symmetric case $y=1/2$, is given  by:
\bea G^{{1\over 2},C}&\equiv& -{1-\gamma^2\over 4}\lim_{\Omega \to 0}\Omega \int_0^\infty dt e^{i(\Omega +i\eta )t}\braket{[N_\psi (t),N_\psi (0)]}_C\br
 &=& -(1-\gamma^2){2e^2\over h}\int d\epsilon_p[-f'(\epsilon_p)]\Delta {\rm T}_{pp}(\epsilon_p)\label{GCyt0}\eea
where $\Delta {\rm T}_{pp}(\epsilon_p)$ is defined in Eq.\,(\ref{eqDeltaT}).  
An important check on this result is that for the non-interacting case, where $G^C$ must be zero, $\Delta {\rm T}_{pp}$ vanishes
due to the optical theorem. (For the interacting case $\Delta {\rm T}_{pp}$ is generally non-zero due to the 
contribution of multi-particle final states to the optical theorem.) 
Unfortunately, for the ABK ring, $\Delta N_{\psi \psi}$ is generally {\it not} a conserved quantity. 

Another special case where the connected part can be eliminated is with parity symmetric  $\gamma =1$, $\varphi =0$.  Now
$\mathbb{U}_k$ becomes the identity matrix since only the even channel couples to the impurity.  Thus:
\be  \mathbb{M}^{y11}_{k_1k_2}=2\pi \delta (k_1-k_2)(2y-1)\ee
which is zero for the parity symmetric choice $y=1/2$.  Then Eqs.\,(\ref{eqzr2}) to (\ref{eqzc2}) reduce to:
\bea 
\mathcal{Z}_R(\epsilon_p)&=&{4\tau \cos p+2(1+\tau^2)\over 1+\tau^2+2\tau \cos p}\sqrt{\mathcal{T}_0[1-\mathcal{T}_0]},\\ 
\mathcal{Z}_I(\epsilon_p)&=&1-2\mathcal{T}_0(\epsilon_p),\vsp\\  
\mathcal{Z}_2(\eps_p)&=&0\vsp.
\eea 
Yet another special case is when $t_R$ or $t_L=0$.  Then we find
\be \mathbb{M}^{y11}_{k_1k_2}=2\pi \delta (k_1-k_2)\left[ 2y-1-{1-\tau^2\over 1+\tau^2}\right]
\ee
which is zero for
\be y={1\over 1+\tau^2}.\label{ytL0}\ee
 Eqs.\,(\ref{eqzr2}), (\ref{eqzi2y}) and (\ref{eqzc2y}) then give, for this value of $y$:
 \bea \mathcal{Z}_R(\epsilon_p)&=&{4\tau \cos p\over 1+\tau^2}\sqrt{\mathcal{T}_0(\epsilon_p)[1-\mathcal{T}_0(\epsilon_p)]}\br
 \mathcal{Z}_I(\epsilon_p)&=&1-2\mathcal{T}_0(\epsilon_p)-\left({1-\tau^2\over 1+\tau^2}\right)^2\vsp\br
 \mathcal{Z}_2(\eps_p)&=&0\vsp.\eea

\noindent
In general, the crucial function $ \mathbb{M}^{y11}_{k_1k_2}$ has the form:
\bea
M^{y11}_{k_1k_2}&=&[2y-1+m_{k_1k_2}^{c}]2\pi\delta(k_1-k_2)\vsp\br &+&2it\sin k_1{m^{o}_{k_1k_2}}g^R_{k_1}(\eps_{k_2})\label{My11}
\eea
where the functions $m_{pp}^{c}$ and $m^{o}_{pp}$, in the diagonal case $k_1=k_2=p$, are given in Eqs.\,(\ref{eqmc}) and (\ref{eqmo}) respectively. 
Clearly, for general values of $\tau$, $\gamma$ and $\varphi$,  we cannot make $M^{y11}_{k_1k_2}$ zero for all $k_1$ and $k_2$ for any choice of $y$. \\

\subsection{Using the ``transmission probability'' expression for the connected and disconnected parts of the conductance}
\label{useTP}
In Sec.\,\pref{sec:cond} we showed that both disconnected and connected parts of the conductance can be written in the form:
\be G^{D/C}={2e^2\over h}\int d\epsilon_p[-f'(\epsilon_p)]\mathcal{T}^{D/C}(\epsilon_p)
\ee
defining disconnected and connected parts of an energy-dependent ``transmission probability''.
It is convenient to define the imaginary frequency Green's function:
\bw

\be Q^{y}_{k_1k_2}(i\varpi_m,i\varpi_m+i\omega_p)\equiv 
 V_{k_1}V_{k_2}\intob{d\tau_1d\tau_2e^{i\varpi_m\tau_1+i(\varpi_m+\omega_p)\tau_2}\braket{T_{\tau}
d\dg(\tau_1)d\dn(\tau_2)N_y(0)}}_C.\ee
Then the function $P(i\varpi_m,i\varpi_m+i\omega_p)$ defined in Eq.\,(\ref{Pdef}), generalized to finite $y$, becomes:
 \bea &&P^y(i\varpi_m,i\varpi_m+i\omega_p)\equiv \frac{1}{4}\intopi{\frac{dk_1dk_2}{(2\pi)^2}} 
M^{y11}_{k_1k_2} g_{k_1}(i\varpi_m)g_{k_2}(i\varpi_m+i\omega_p)\mathcal{Q}^{y}_{k_1k_2}(i\varpi_m,i\varpi_m+i\omega_p).
\label{Pydef}
\eea
 The connected part of the transmission probability can now be written:
 \bea \mathcal{T}^{yC}(\omega )&=&\lim_{\Omega \to 0}{\Omega^2\over 4}P^y(\omega -i\eta_1,\omega +\Omega 
+i\eta_2)+c.c. \nonumber \\ 
&=&\lim_{\Omega \to 0}{\Omega^2\over 16}\intopi{\frac{dk_1dk_2}{(2\pi)^2}} 
M^{y11}_{k_1k_2} g^A_{k_1}(\omega )g^R_{k_2}(\omega +\Omega )
\cdot Q^{y}_{k_1k_2}(\omega -i\eta_1,\omega +\Omega 
+i\eta_2)+c.c.
\eea
where $Q^{y}_{k_1k_2}(\omega -i\eta_1,\omega +\Omega +i\eta_2)$ is the analytic continuation of 
$Q^{y}_{k_1k_2}(i\varpi_m,i\varpi_m+i\omega_p)$ to real frequencies. 
 Using the expression for $M^{y11}_{k_1k_2}$ in Eq.\,(\ref{My11}) and the propagator product identities of App.\,(\ref{sec:ppi}) we see that
 \be \lim_{\Omega \to 0}\Omega M^{y11}_{k_1k_2} g^A_{k_1}(\epsilon_p )g^R_{k_2}(\epsilon_p +\Omega )
 =2\pi \nu_p[2y-1+m^c_{pp}+m^o_{pp}]\delta (k_1-k_2)\delta (k_1-p)
 .\label{ppic}
 \ee 

\ew

We expect an additional factor of $1/\Omega$ to arise from $Q^{y}_{k_1k_2}(\omega -i\eta_1,\omega +\Omega +i\eta_2)$ as $\Omega \to 0$. Thus we may write:
 \be \mathcal{T}^{yC}(\epsilon_p)=-[2y-1+m^c_{pp}+m^o_{pp}]F_y(\epsilon_p)\ee
 where
\be F_y(\epsilon_p)\equiv {\pi \nu_p\over 8} \lim_{\Omega \to 0}\Omega Q^{y}_{pp}(\epsilon_p -i\eta_1,\epsilon_p +\Omega +i\eta_2)
 +c.c.\nonumber\ee
Due to the $N_y$ operator in $Q^y$, $F_y(\epsilon_p)$ is a sum of terms of zeroth and first order in $y$.
Despite the simplifications resulting from using the transmission probability expression for the conductance, it appears that no choice of 
 $y$ will make the connected part vanish.  Nonetheless, the fact that the total conductance must be independent of $y$ implies 
 some relationship between $\mathcal{T}^{yC}(\epsilon_p)$ and $\mathcal{T}^{yD}(\epsilon_p)$. This implied relationship  involves
 the integral of these functions.  
However it is interesting to consider the consequences of the stronger assumption that 
$\mathcal{T}^{yC}(\epsilon_p)+\mathcal{T}^{yD}(\epsilon_p)$ is independent of $y$ for all energies.  Using the expression for $\mathcal{T}^{yD}(\epsilon_p)$ 
in Eqs.\,(\ref{eqzc2y}) to (\ref{eqDeltaT}), we may write:
\bw

\bea 
&&\mathcal{T}^{yC}(\epsilon_p)+\mathcal{T}^{yD}(\epsilon_p)=\mathcal{T}(\epsilon_p)+(2y-1+m^c_{pp}+m^o_{pp})[(2y-1+m^c_{pp})\Delta {\rm T}_{pp}-F_y(\epsilon_p)]\qquad
\eea
$\Delta {\rm T}_{pp}$, given in Eq.\,(\ref{eqDeltaT}), measures violations of the optical theorem when the ${\rm T}$-matrix is 
 restricted to the single particle sector. 
 As mentioned above, $F_y(\epsilon_p)$ is a sum of terms of zeroth and first order in $y$. We see that $y$-independence of 
 $\mathcal{T}^{yC}(\epsilon_p)+\mathcal{T}^{yD}(\epsilon_p)$ would require:
 \be F_y(\epsilon_p)=(2y-1+m^c_{pp})\Delta {\rm T}_{pp}\ ??\ee
Then, the total transmission probability becomes
 ${\cal T}(\epsilon_p)$, given in Eq.\,(\ref{Tdef}). 
 It can be seen that our perturbative calculation is actually consistent with this stronger assumption. In this case
\bea
F_y(\eps_p)&=&\frac{\pi\nu_p}{2}\im{\lim_{\Omega\ra 0}\Omega\intopi{\frac{dq_1dq_2}{(2\pi)^2}}\frac{3}{16}J_{q_2p}J_{pq_1}\Big( M_{q_1q_2}^{y11}g^R_{q_1}(\eps_p+\Omega)g^A_{q_2}(\eps_p)\Big)}\br
&=&[2y-1+m_{pp}^{c}]\frac{3\pi^2}{16}\nu_p^2J_{pp}^2=[2y-1+m_{pp}^{c}]\Delta {\rm T}_{pp}\label{Fpert}
\eea
\ew

where the fact that ${\rm T}_{pp}$ is $O(J^2)$ was used in the last step so that $\Delta {\rm T}$ reduces to Im$[-\pi \nu_p{\rm T}_{pp}(\epsilon_p)]$, 
whose perturbative value is given in Eq.\,(\ref{eqtmatrix}). However, it seems unlikely that this stronger assumption will  survive higher orders of perturbation theory 
so we now proceed without making it.

 \subsection{Approximate elimination of connected term}
\label{approx}
 Since the connected part of the conductance is given in terms of $\mathcal{T}^{yC}(\epsilon_p)$ by Eq.\,(\ref{GCT}),  we see that, for 
 temperatures small compared to the band width, $G^{yC}$ will approximately vanish provided we choose $y$ so that  $\mathcal{T}^{yC}(\epsilon_p)$ 
 vanishes at the Fermi energy, $\epsilon_p=\epsilon_F$. This choice is:
 \be y={1\over 2}[1-m^c_{p_Fp_F}-m^o_{p_Fp_F}].\label{ysp}\ee
  Since $m^o_{pp}+m^c_{pp}$ is a  smooth function of $\epsilon_p$ as can 
be seen from Eqs.\,(\ref{eqmo}) and (\ref{eqmc}), we may calculate the leading contribution of $G^{yC}$ to the conductance, for this choice of $y$, by the Sommerfeld 
 expansion, giving a suppression factor of order $(T/t)^2$ where $T$ is the temperature and $4t$ the band width.
We see from Eq.\,(\ref{eqzi2y}) that ${\cal Z}_2^y(\epsilon_F)$ vanishes for the special value of $y$, Eq.\,(\ref{ysp}) which makes the connected part approximately vanish. 
The reason for this can be seen in App.\,(\ref{sec:ap02}).  The same product of $M^{y11}_{k_1k_2}$ and propagators occurs in Eq.\,(\ref{eqID2}) 
for the term quadratic in the ${\rm T}$-matrix 
as in Eq.\,(\ref{ppic}) for the connected part.
Thus we may write the conductance for $T\ll t$ as a linear function of the ${\rm T}$-matrix:
\bea
G\approx {2e^2\over h}\int{d\eps_p [-f'(\eps_p)]}&\Big\{&
\mathcal{T}_0(\eps_F)\br
&&\hspace{-0.8cm}+\mathcal{Z}_R(\eps_F)\re{ -\pi\nu_{p}{\rm T}_{pp}(\eps_p)}\vsp\br 
&&\hspace{-0.8cm}+\mathcal{Z}_I'(\eps_F)\im{-\pi\nu_{p}{\rm T}_{pp}(\eps_p)}\Big\}\qquad\label{G}
\eea
where $\mathcal{T}_0(\eps_p)$,  $\mathcal{Z}_R(\eps_p)$ and $\mathcal{Z}_I'(\eps_p)$ are given in Eqs.\,(\ref{eqz02}), (\ref{eqzr2}) and (\ref{Z'def}) 
respectively. 
Eq.\,(\ref{G}), along with our formulas for the coefficients, is one of the main results of this paper.  It shows that the conductance through the small ABK ring can be expressed entirely in terms of the ${\rm T}$-matrix of a single channel Kondo or Anderson model at temperatures small compared to the band width. Note that within second order perturbation theory in Kondo coupling (valid at $T\gg T_K$), the T-matrix is also smooth and can be approximated by its value at the Fermi and taken out of the integral. However, at lower temperatures the T-matrix contains sharp features on the scale of $T_K$ and the thermal averaging is relevant.

It is interesting to compare this result to  [\onlinecite{ Hofstetter}] which also gives a formula for the conductance of the short ABK ring as a sum of terms 
of zeroth and first order in the ${\rm T}$-matrix.  A precise agreement cannot be expected since  [\onlinecite{ Hofstetter}] assumes energy 
independent tunneling parameters and expresses the result in terms of parameters at the Fermi surface only. We find precise 
agreement at half-filling, $p_F=\pi /2$, only.  

Note that our argument depends crucially on the fact that   $m^o_{pp}$ and $m^c_{pp}$ defined in Eqs.\,(\ref{eqmo}) and (\ref{eqmc}), are smooth functions of $\epsilon_p$; the energy scale over which they vary significantly is the band width, $4t$. 
However, we expect this not to be the characteristic energy scale for a large ring of length $L$. 
The problem is that  a small energy scale enters the calculation, 
the finite size gap $\propto v_F/L$.  We then expect the analogue of  $m^o_{pp}+m^c_{pp}$
 to vary on this scale, making the approximate elimination of $G^C$ only possible for $T\ll v_F/L$ which is much less than the band width 
 for a ring much larger than a lattice constant. Thus, we may expect that a calculation of the connected part of the conductance will be necessary, 
 except at extremely low temperatures.

\subsection{Keldysh approach}\label{sec:Keldysh}
The presence of the connected four-point function in the conductance (even if it can be approximately eliminated) is surprising given that, according to [\onlinecite{Hofstetter}] and following the Keldysh approach of Meir and Wingreen [\onlinecite{Meir}], the conductance can be expressed entirely in terms of the retarded two-point function $G_{dd}^R$. In this sub-section, we calculate the conductance using Keldysh approach and point out that generally symmetrization fails and the equilibrium two-point Green's functions are not sufficient for determining the conductance. However, similar to the discussion of Kubo section, at temperatures small compared to the band width the non-equilibrium Green's functions can be approximately eliminated. Similar calculations have been reported previously in [\onlinecite{Bruder}-\onlinecite{Hofstetter},\onlinecite{Ueda}]. Going back to the Anderson model and defining $N_L=\sum_{n>0}{c\dg_{-n}c\dn_{-n}}$, we can write the current in the left lead as
\bea\label{eqIdefK}
{I}_L&=&\frac{ie}{\hbar}\braket{\com{\hat N_L}{\hat H}}\br &=&-\frac{2e}{\hbar}\re{t_Le^{i\varphi/2}{\nq G^<_{dL}(0)}+t'{\nq G^<_{RL}(0)}}
\eea
which is expressed in terms of \emph{non-equilibrium} equal-time lesser Green's functions defined as $\nq G_{dL}^<(t)=i\langle{\hat c\dg_{-1}(t)\hat d(0)}\rangle$ and 
$\nq G_{RL}^<(t)=
i\langle{\hat c\dg_{-1}(t)\hat c\dn_{1}(0)}\rangle$ involving the dot and the first sites of the left and right leads. Here $L$ and $R$ indices represent sites -1 and 1 respectively. We denote non-equilibrium operators with a hat and corresponding Green's functions with Sans-serif font. Replacing equal-time Green's functions with a frequency integral over the Fourier transform of corresponding unequal-time Green's functions we get
\be
\hspace{-.15cm}I_L=\intinf{d\omega}I_L(\omega), \quad I_L(\omega)=-\frac{2e}{h}\re{\nq 
W_L^<(\omega)}
\ee
where
\be
 \nq W_L^<(\omega)\equiv 2t_Le^{i\varphi/2}{\nq G^<_{dL}(\omega)}+2t'{\nq G^<_{RL}(\omega)}.
\ee
In the non-interacting case, $I_L(\omega)$ can be interpreted as the contribution to the current from electrons of energy $\omega$.
Following Meir-Wingreen the mixed functions $\nq G_{dL}$ and $\nq G_{RL}$ are related to the Green's function of the dot. For that purpose, the Green's functions are generalized to complex times
on Keldysh contour $C_K$ and we use the equation of motion to obtain
\bea
\nq G_{dL}(\tau,\tau')=&-&t_Le^{-i\varphi/2}\int_{C_K}{d\tau_1 \nq G_{dd}(\tau,\tau_1) g_{LL}(\tau_1,\tau')}\br &-&t'\int_{C_K}{d\tau_1}\nq G_{dR}(\tau,\tau_1) g_{LL}(\tau_1,\tau').
\eea
Here $\nq{G}_{dL}(\tau,\tau')\equiv\langle-iT_C d\dn(\tau)c_{-1}\dg(\tau')\rangle$ is the contour-ordered mixed Green's function and $g_{LL}(\tau,\tau')$ is the Green's function of the first site of the decoupled left lead in \emph{equilibrium} with its own electrochemical potential $\mu_L$.
Going to real time and taking the Fourier transform we can represent this equation in Keldysh space by the matrix equation [\onlinecite{Rammer}]
\be
\breve{\nq G}_{dL}(\omega)=-t_Le^{-i\varphi/2}\breve{\nq G}_{dd}(\omega)\breve{g}_{LL}(\omega)-t' \breve{\nq G}_{dR}(\omega)\breve{g}_{LL}(\omega).\label{eqgdl2}
\ee
Here the $\bnq G(\omega)$ and $\breve {g}(\omega)$'s are $2\times 2$ matrices in Keldysh space whose structure is
\be
\bnq G(\omega)=\mat{\nq G^R(\omega)& \nq G^K(\omega) \\ 0 & \nq G^A(\omega)},
\ee
with $\nq G^K(\omega)\equiv \nq G^>(\omega)+\nq G^<(\omega)$. 
Similarly, for the other Green's functions, suppressing energy-dependences, we can write
\bea
\breve{\nq G}_{dR}&=&-t_Re^{i\varphi/2}\breve{\nq G}_{dd}\breve{g}_{RR}-t' \breve{\nq G}_{dL}\breve{g}_{RR}\label{eqgdr2}\vsp\\
\breve{\nq G}_{LL}&=&\breve{g}_{LL}-t_Le^{i\varphi/2}\breve{g}_{LL}\breve{\nq G}_{dL}-t' \breve{g}_{LL}\breve{G}_{RL}\label{eqgll2}\vsp\\
\breve{\nq G}_{RL}&=&-t_R e^{-i\varphi/2}\breve{g}_{RR}\breve{\nq G}_{dL}-t' \breve{g}_{RR}\breve{\nq G}_{LL}. \vsp \label{eqgrl2}
\eea
These equations can be alternatively derived starting from a representation of the Hamiltonian in momentum space. 
After some algebra we arrive at the matrix equation
\bea\label{eqxwx}
\breve{X}_L\bnq{W}_L\breve{X}_L=&-&2t'^2\breve{g}_R\breve{g}_L\breve{X}_L+2\breve{P}_L\bnq{G}_{dd}\breve{Q}_L
\eea
where $\breve{X}_L(\omega)$, $\breve{P}_L(\omega)$, and $\breve{Q}_L(\omega)$ are given by
\bea\label{eqxwx2}
\breve{X}_L&=&\bv{1}-t'^2\breve{g}_R\breve{g}_L,\vsp\br
\breve{P}_L&=&\bv{1}t_Le^{i\varphi/2}-\breve{g}_Rt't_Re^{-i\varphi/2},\vsp\br
\breve{Q}_L&=&-t_Le^{-i\varphi/2}\breve{g}_L+t't_Re^{i\varphi/2}\breve{g}_R\breve{g}_L.\vsp
\eea
The real part of the lesser component of $\bnq W_L(\omega)$ gives the energy-resolved current from the left lead $I_L(\omega)$
\bea
\frac{h}{2e}I_L(\omega,\Delta\mu)&=&z^{\rm KL}_{0}(\omega,\Delta\mu)\br
&+&z_{R}^{\rm KL}(\omega,\Delta\mu)\re{\nq G_{dd}^R(\omega)}\vsp\br 
&+&z_{I}^{\rm KL}(\omega,\Delta\mu)\im{\nq G_{dd}^R(\omega)}\vsp\br 
&+&z_<^{\rm KL}(\omega,\Delta\mu)\left[i\nq G_{dd}^<(\omega)\vsp   \right].
\eea
The subscripts of the $z^{\rm KL}$ coefficients are related to the corresponding correlation function of the dot and their superscript means they are obtained from Keldysh technique and related to the left lead. The current in the right lead is obtained from $L\lr R$ and $\varphi\lr-\varphi$ substitution. Assuming a symmetric applied bias, the bias dependence of the $z$ coefficients are caused by $f_{L/R}(\omega)=f(\omega)\pm\frac{1}{2}\Delta\mu f'(\omega)+\cdots$ inside $\breve g_{L}(\omega)$ and $\breve g_R(\omega)$ and are indicated explicitly, but the Green's functions also have an implied bias-dependence. Using Eq.\,\pref{eqxwx}, it can be shown that the $z$ coefficients have a Taylor series in the bias $\Delta\mu$ of the form
\bea\label{eqztaylor}
\vsp z_0^{\rm KL}(\omega,\Delta\mu)&=&\Delta\mu[-f'(\omega)]\mathcal{T}_0(\omega)+\cdots,\br
\vsp z_I^{\rm KL}(\omega,\Delta\mu)&=&z_I^{\rm KL}(\omega,0)+\Delta\mu[-f'(\omega)]\mathcal{Z}_I^{\rm KL}(\omega)+\cdots,\br
\vsp z_R^{\rm KL}(\omega,\Delta\mu)&=&\Delta\mu[-f'(\omega)]\mathcal{Z}_R^{\rm KL}(\omega)+\cdots,
\eea
defining the coefficients $\mathcal{Z}^{\rm KL}(\omega)$, whereas $z_<(\omega,\Delta\mu)=z_<(\omega)$ is independent of the bias. The equilibrium components of the first two terms are zero, $z_0^{\rm KL}(\omega,0)=z_R^{\rm KL}(\omega,0)=0$. But $z_I(\omega,0)$ is nonzero and satisfies~[\onlinecite{Meir}]
\be\label{eqzizless}
z_I^{\rm KL}(\omega,0)=-2{f}(\omega)z_<^{\rm KL}(\omega),
\ee
which by the equilibrium condition $iG^<_{dd}(\omega)=2f(\omega)\im{G^R_{dd}(\omega)}$ ensures that the expectation-value of the current operator defined by Eq.\,\pref{eqIdefK} is indeed zero in equilibrium. The function $\mathcal{T}_0(\omega)$ is the same background transmission we had in Kubo calculations. Using Eqs.\,\pref{eqztaylor}-\pref{eqzizless} we can write the current at energy $\omega=\eps_p$ as
\bea\label{eqIdelmu}
\frac{h}{2e}I_L(\eps_p,\Delta\mu)&=&\Delta\mu[-f'(\eps_p)]\Big\{\mathcal{T}_0(\eps_p)\vsp\br &+&\mathcal{Z}^{\rm KL}_R(\eps_p)\re{G^R_{dd}(\eps_p)}\vsp\br &+&\mathcal{Z}^{\rm KL}_I(\eps_p)\im{G_{dd}^R(\eps_p)}\Big\}+O(\Delta\mu^2)\vsp\br
&+&z_<^{\rm KL}(\eps_p)\Pi(\eps_p,\Delta\mu),
\eea
where 
\be
\Pi(\eps_p,\Delta\mu)\equiv i{\nq G}^<_{dd}(\eps_p)-2f(\eps_p)\im{{\nq G}_{dd}^R(\eps_p)}.
\ee
The first four lines of Eq.\,\pref{eqIdelmu} contains two-point equilibrium Green's function of the dot whereas the last line, written in terms of $\Pi(\omega,\Delta\mu)$ contains non-equilibrium Green's functions and is more complicated to compute. Since $z_<^{\rm KL}(\omega)$ is nonzero, in order to get the linear-response current in general one needs to do a first-order perturbative-in-bias expansion of the non-equilibrium functions in $\Pi(\omega,\Delta\mu)$, which leads to both connected four-point and disconnected two-point contributions related to the terms proportional to $\mathcal{Z}_2^y$ in the Kubo framework. 

The Meir-Wingreen approach [\onlinecite{Meir}] uses the fact that the DC current satisfies $\langle\hat I_L+\hat I_R\rangle=0$ to symmetrize the current between left and right leads in order to eliminate the non-equilibrium Green's function of the dot [second line of Eq.\,\pref{eqIdelmu}] for the case of no reference arm. That would mean that the linear conductance of the system is entirely given by the \emph{equilibrium} two-point function $G_{dd}^R(\omega)$.
However, such a procedure fails for the present problem as already noticed by Dinu et al. [\onlinecite{Dinu}]. This can be seen most easily by taking the three sites $-1$, $d$ and $1$ as the central sites of the device and noticing that the coupling matrices introduced by Meir-Wingreen [\onlinecite{Meir}] do not satisfy their ``proportional coupling'' condition. Equivalently, one can attempt to find a parameter $y$ for which the symmetrized current $I_y=yI_R-(1-y)I_L$ is not a functional of $\Pi(\omega,\Delta\mu)$. Generally we can use
\bea\label{eqyepsk}
y(\eps_F)&\equiv&\frac{z^{\rm KL}_<(\eps_F)}{z^{\rm KL}_<(\eps_F)+z^{\rm KR}_<(\eps_F)}\br&=&\frac{(t_L^2+t_R^2\tau^2)+2t_Lt_R\tau\cos(p_F-\varphi)}{(t_L^2+t_R^2)(1+\tau^2)+4t_Lt_R\tau\cos p_F\cos\varphi}.\qquad
\eea
to eliminate the non-equilibrium functions at the Fermi energy.
This is precisely the same condition on $y$ obtained in Eq.\,(\ref{ysp}) using the Kubo approach. 
This function is independent of energy if and only if at least one of the three parameters $t_L$, $t_R$ or $\tau=t'/t$ is zero or for the case when 
$t_L=t_R$ and $\varphi=0$, indicating that the total symmetrized current does not contain $\Pi(\eps_p,\Delta\mu)$ in these special cases. These are precisely the special cases discussed above using the Kubo approach. The energy dependence of $y(\omega)$ for general parameters indicates that non-trivial symmetrization requires $I_L(\omega)+I_R(\omega)$ to be zero (conservation of energy-resolved currents) which is not the case in interacting systems at finite temperature. However, it is expected that the function $\Pi(\omega,\Delta\mu)$ contains a derivative of the Fermi distribution function, i.e. we can write
\be
\Pi(\eps_p,\Delta\mu)=\Delta\mu[-f'(\eps_p)]\Pi'(\eps_p)+O(\Delta\mu^2).
\ee
Therefore, for temperatures smaller than the band width, the slowly varying parameter $y(\eps_p)\approx y(\eps_F)$ is constant within $I_L$ the energy integral range and can be taken out of the integral and again an \emph{approximate} symmetrization can be used to eliminate the difficult function $\Pi'(\eps_p)$ by setting its coefficient approximately equal to zero. The parameter $\mathcal{Z}^{\rm KL}_R(\eps_p)$ is the same for both leads and are unaffected by symmetrization and it is $-\pi\nu_pV_p^2$ times the corresponding parameter in the Kubo calculations. The parameters $\mathcal{Z}_I^{\rm KL}(\eps_p)$ and $\mathcal{Z}_I^{\rm KR}(\eps_p)$ are, however, different but the symmetrized parameter $y(\eps_F)\mathcal{Z}_I^{\rm KR}(\eps_p)+[1-y(\eps_F)]\mathcal{Z}_I^{\rm KL}(\eps_p)$ is equal $-\pi\nu_pV_p^2$ times the parameter 
$\mathcal{Z}_I{'}(\eps_p)$ obtained in the Kubo section at $\eps_p=\eps_F$.

\section{Conclusions}
\label{conc}
We have studied transport properties of a small Aharonov-Bohm ring with an embedded quantum dot in one of its arms. The DC conductance is calculated using the Kubo formula and it is shown that there is a contribution which involves a connected four-point function. We have shown that for $T$ small compared to the band width, this term and terms quadratic in the T-matrix can be eliminated, leaving a formula 
for the conductance linear in the T-matrix. This is a useful result because rather precise results exist on the T-matrix 
for a wide range of temperatures and frequencies, using renormalization group improved perturbation theory, Nozi\' eres Fermi 
liquid theory, numerical renormalization group and other methods. 
We have calculated the conductance 
perturbatively in the Kondo coupling, a result that should be valid for $T\gg T_K$.

A natural question to ask is whether our $O(J^2)$ results are consistent with the observation that Kondo scattering is largely 
inelastic at  $T\gg T_K$ [\onlinecite{Zarand04}].  This observation simply follows from the fact that the (single-particle) ${\rm T}$-matrix of the Kondo model 
starts with an imaginary term of order $J^2$. Then the optical theorem:
\be\label{tmany} 
-i(\hat {\rm T}-\hat {\rm T}\dg)=\hat{\rm T}\dg \hat{\rm T}
\ee
is badly violated {\it if the sum over intermediate states, inserted between $\hat{\rm T}\dg$ and $\hat{\rm T}$  is restricted to the single-particle sector} [\onlinecite{Zarand04}]. 
The full flux dependence of the conductance through an ABK ring is complicated. Using our approach, it arises 
partly from the flux dependence of the coupling of the quantum dot to the screening channel, which introduces 
a flux dependence of the Kondo coupling and hence the Kondo temperature. Further flux dependence arises from 
the $\mathcal{Z}_R(\eps_p)$ and $\mathcal{Z}_I'(\eps_p)$ coefficients given in Eqs.\,\pref{eqzr2} and \pref{Z'def} relating the conductance to the real and imaginary parts of the T-matrix. (At higher temperatures, where the connected part must be included, the flux dependence becomes even more complicated.) 
The conductance is quadratic in the Kondo coupling, in perturbation theory, while being first order in the 
potential scattering. Thus, the absence of a term linear in the Kondo coupling leads to a reduction of 
flux dependence at high $T$ and can be ``explained'' by the fact that the scattering is purely inelastic in that 
limit [\onlinecite{Carmi}].

We leave the extension of our results to lower temperature and to larger rings for future work [\onlinecite{KZA}]. 
As discussed in Sec.\,\pref{sec:con}, for a large ring of length $L$ the connected term in the conductance can only be safely eliminated at temperatures 
below the finite size energy level spacing ($T\ll v_F/L$). 
Thus, a thorough treatment will probably require calculation of the novel connected 4-point Green's function at lower temperatures. 
Moreover, the relation between the degree of flux dependence of the conductance and the degree of inelastic scattering in general also remains an open question. 
\subsection*{Acknowledgement}
\noindent
We thank M. Eto and Z. Shi for stimulating discussions. This work was supported by NSERC and CIfAR. Y.~K.~gratefully acknowledges financial support from the Swiss National Science Foundation. R.~Y.~is the Yukawa Fellow and this work is partially supported by Yukawa Memorial Foundation.

\appendix
\section{Propagator product identities}
\label{sec:ppi}
In this Appendix we show explicitly how the limit $\Omega \to 0$ is taken in various equations in this paper.  The results presented in this Appendix also 
support our argument that the first and last terms in Eq.\,(\ref{intP}) can be dropped as $\Omega \to 0$. 
We start by considering $\lim_{\Omega \to 0}\Omega g^R_k(\omega )g^A_k(\omega +\Omega )$. 
We use:
\bea 
{1\over \omega -\epsilon_k+i\eta}-{1\over \omega +\Omega -\epsilon_k-i\eta}
 =\br
&&\hspace{-2cm} {\Omega -2i\eta \over (\omega -\epsilon_k+i\eta )(\omega +\Omega -\epsilon_k-i\eta )}.\qquad
 \eea\\
 
\noindent 
Thus
\be \lim_{\Omega \to 0}\Omega g^R_k(\omega )g^A_k(\omega +\Omega )=g^R_k(\omega )-g^A_k(\omega )=-2\pi i\delta (\omega -\epsilon_k)
\label{PPE1}.\ee
On the other hand, the same reasoning gives
\be \lim_{\Omega \to 0}\Omega g^R_k(\omega )g^R_k(\omega +\Omega )=0.\ee
These results can be checked by doing the $\omega$ integral. 
\be \int _{-\infty}^\infty d\omega {1\over (\omega +i\eta )(\omega +\Omega -i\eta )}={2\pi i\over -\Omega }.\ee
The integral is non-zero since the poles are on opposite sides of the real $\omega$ axis. On the other 
hand the integral of $g^R_k(\omega )g^R_k(\omega +\Omega )$ is zero because both poles are on the same side. 

The other important propagator product identity involves 3 propagators, $\lim_{\Omega \to 0}\Omega g^A_{k_1}(\omega )
g^R_{k_2}(\omega +\Omega )g^R_{k_1}(\epsilon_{k_2})$. Now we use:
\bw

\bea &&\left[ {1\over \omega -\epsilon_1-i\eta_1}-{1\over \omega +\Omega -\epsilon_2+i\eta_2}\right]
\left[{1\over \epsilon_2-\epsilon_1-\Omega -i\eta_3}-{1\over \epsilon_2-\epsilon_1+i\eta_4}\right]\nonumber \\ &&
\hspace{4cm}=-{\Omega +i(\eta_3+\eta_4)\over (\omega -\epsilon_1-i\eta_1)( \omega +\Omega -\epsilon_2+i\eta_2)( \epsilon_2-\epsilon_1+i\eta_4)}
 \cdot {\epsilon_2-\epsilon_1-\Omega -i(\eta_1+\eta_2)\over \epsilon_2-\epsilon_1-\Omega -i\eta_3}.
\eea
Thus, we conclude
\be \lim_{\Omega \to 0} \Omega g^A_{k_1}(\omega )
g^R_{k_2}(\omega +\Omega )g^R_{k_1}(\epsilon_{k_2})=(2\pi )^2\delta (\omega -\epsilon_{k_1})
\delta (\epsilon_{k_1}-\epsilon_{k_2}).\label{PPE2}\ee
Again this result can be checked by doing the $\omega$ integral:
\be  \Omega \int_{-\infty}^\infty d\omega {1\over( \omega -\epsilon_1-i\eta )
(\omega +\Omega -\epsilon_2+i\eta )(\epsilon_2-\epsilon_1+i\eta )}
=-2\pi  i{\Omega \over (\epsilon_1-\epsilon_2+\Omega +2i\eta )(\epsilon_1-\epsilon_2-i\eta )}
.\label{int:PPE}\ee

\ew

The result of the previous paragraph tells us that the limit $\Omega \to 0$ of the right hand 
side is $(2\pi)^2\delta (\epsilon_{1}-\epsilon_{2})$, consistent with Eq.\,(\ref{PPE2}).
Note that, again, it is crucial which side of the real axis the poles are on. It can be 
seen from Eq.\,(\ref{int:PPE}) that complex conjugating any one of the propagators 
gives zero as $\Omega \to 0$. It is also important to note that
\be g^R_{k_1}(\epsilon_{k_2})=-g^A_{k_2}(\epsilon_{k_1}),\ee
allowing the identity in Eq.\,(\ref{PPE2}) to be written in another equivalent way.

\section{Disconnected contribution to the transmission probability}\label{sec:ap02}
Here we present the details that lead to Eqs.\,(\ref{eqTDanderson})-(\ref{eqzc2}) for the disconnected 
part of the transmission probability, starting from Eqs.\,(\ref{trpr}) and (\ref{eqgTg}). 
From Eq.\,(\ref{eqgTg}) we obtain
\bea -2\hbox{Im}G^R_{q_2 k_1}(\eps_p)&=&(2\pi)^2\delta(q_2-k_1){\delta(\eps_p-\eps_{k_1})}\mathbb{1}\br &&\hspace{-1cm}+\tau_{\psi}\Big(
g^R_{q_2}(\eps_p){V}_{q_2}
i{G}_{dd}^R(\eps_p)
{V}_{k_1}g^R_{k_1}(\eps_p)\br &&-
g^A_{q_2}(\eps_p){V}_{q_2}
i{G}_{dd}^A(\eps_p)
{V}_{k_1}g^A_{k_1}(\eps_p)\Big).\qquad \label{eqG2p}
\eea
 We denote the first term of this expression by a subscript $0$, the second by $R$ and the third by $A$.
Plugging these into Eq.\,\pref{trpr} we get 9 terms for the disconnected part of the transmission probability $\mathcal{T}^D$, 
which we label $\mathcal{T}^D_{00}$, $\mathcal{T}^D_{0R}$, $\mathcal{T}^D_{0A}\cdots\mathcal{T}^D_{AA}$. In the following we write 
\be
\eps_{p'}\equiv-2t\cos p'\equiv\eps_p+\Omega,
\ee
defining a (positive) momentum $p'$ in terms of $p$ and $\Omega$.
\subsection{Background transmission probability}
\noindent
The background transmission probability $\mathcal{T}_0(\eps_p)\equiv\mathcal{T}_{00}^D(\eps_p)$ is obtained by 
choosing the  delta-function term from Eq.\,(\ref{eqG2p}), in both factors of $\hbox{Im}G^R$ in Eq.\,(\ref{trpr}) which leads to
\bea
\mathcal{T}_{0}(\eps_p)&=&
\frac{1}{8}\lim\limits_{\Omega\ra 0}\Omega^2 {(2\pi)^2}\nu_p\nu_{p'}\frac{1}{\Omega+i\eta}\frac{1}{-\Omega+i\eta}\tr{\mathbb{F}_{pp'}\mathbb{F}_{p'p}}
\br &=&\frac{4\tau^2\sin^2p}{(1+\tau^2)^2-4\tau^2\cos^2p}
\label{eqz01}
\eea
and is a Landauer-type formula for the transmission probability through the reference arm. This part of the conductance comes from the free part of the Green's functions and the non-diagonal (in momentum) part of $\Delta N$ corresponding to the overlap (second) term in $\mathbb{A}_{k_1k_2}$ in Eq.\pref{eqdefa}. The contact term of $\mathbb{A}_{pp'}$ does not contribute since $p\neq p'$, 
 $\Omega >0$ until the end of the calculation.

\subsection{Terms linear in T-matrix}
\noindent
The terms in the transmission probability linear in $G_{dd}$ are 
$\mathcal{T}^{D1}\equiv\mathcal{T}^D_{0R}+\mathcal{T}^D_{R0}+\mathcal{T}^D_{0A}+\mathcal{T}^D_{A0}$
and we only need to calculate the first two as they are the complex conjugate of the second two. Using Eqs.\pref{eqMy} 
and \pref{eqG2p}  the first term is
\bea
\mathcal{T}^D_{0R}(\eps_p)=\frac{1}{8}\lim\limits_{\Omega\ra 0}{2\pi\Omega^2}\nu_{p'}\intopi{
\frac{dk_1dq_2}{(2\pi)^2}}i{G}^R_{dd}(\eps_p)\br 
\tr{\mathbb{U}_{k_1}\mathbb{A}_{k_1p'}\mathbb{A}_{p'q_2}\mathbb{U}\dg_{q_2}g^R_{q_2}(\eps_p){V_{q_2}}\tau_{\psi}{V_{k_1}}g^R_{k_1}(\eps_p)}.
\eea
Using the cyclic property of the trace, this can be re-arranged to give
\be
\mathcal{T}^D_{0R}(\eps_p)=\frac{\pi\nu_p}{4}\lim_{\Omega\ra 0}\tr{\mathbb{I}\dn_{D1}(\eps_p,\Omega)\tau_{\psi}\mathbb{I}'_{D1}(\eps_p,\Omega)}iG_{dd}^R(\eps_p).
\label{trace}\ee
where the matrices $\mathbb{I}_{D1}$ and $\mathbb{I}'_{D1}$ are defined as
\bea\label{eqId1}
\mathbb{I}_{D1}\dn(\eps_p,\Omega)&\equiv&\Omega\intopi{\frac{dq_2}{2\pi}{V_{q_2}}g^R_{q_2}(\eps_p)\mathbb{A}_{p'q_2}\mathbb{U}\dg_{q_2}},\br
\mathbb{I}'_{D1}(\eps_p,\Omega)&\equiv&\Omega\intopi{\frac{dk_1}{2\pi}{ V_{k_1}}g^R_{k_1}(\eps_p)\mathbb{U}_{k_1}}\mathbb{A}_{k_1p'}.
\eea
Using the definition of Eq.\,\pref{eqdefa}, the first one of these can be written as
\bea
&&\mathbb{I}_{D1}\dn(\eps_p,\Omega)=-\tau_x V_{p'}\mathbb{U}\dg_{p'}\br
&&\hspace{1.5cm}-\Omega\intopi{\frac{dq_2}{2\pi}}g^R_{q_2}(\eps_p)g^A_{q_2}(\eps_{p}+\Omega)
\mathbb{F}_{p'q_2}V_{q_2}\mathbb{U}\dg_{q_2}.\qquad
\eea
Using the propogator product identity of Eq.\,(\ref{PPE1}) we obtain 
\bea\label{eqID1}
\mathbb{I}_{D1}\dn(\eps_p,\Omega)\ra \Big(-\tau_x+{2\pi i}\nu_p\mathbb{F}_{pp}\Big)V_p\mathbb{U}\dg_p.
\eea
 Similarly,
\bea\label{eqID1p}
\mathbb{I}'_{D1}(\eps_p,\Omega)\ra-V_{p}\mathbb{U}_{p}\tau_x.
\eea
The reason the $\mathbb{F}$-term is present in Eq.\,\pref{eqID1} but absent in Eq.\,\pref{eqID1p} is that $\mathbb{I}_{D1}$ includes a term $\propto g^R_{q_2}(\eps_p)g^A_{q_2}(\eps_p+\Omega)$  whereas $\mathbb{I}_{D1}'$ includes a term 
$\propto g^R_{k_1}(\eps_p)g^R_{k_1}(\eps_p+\Omega)$.
In terms of $\mathbb{I}_{D1}\dn$ and $\mathbb{I}_{D1}'$  we can write the second contribution to $\mathcal{T}^{D1}$ as
\bea
&&\mathcal{T}^D_{R0}(\eps_p)=\frac{\pi\nu_{p'}}{4}\lim_{\Omega\ra 0}iG_{dd}^R(\eps_{p'})\br
&&\hspace{1.7cm}\tr{\mathbb{I}_{D1}\dn(\eps_p+\Omega,-\Omega)\tau_{\psi}\mathbb{I}'_{D1}(\eps_p+\Omega,-\Omega}.\qquad
\eea
The DC limit of the matrices $\mathbb{I}_{D1}\dn(\eps_p+\Omega,-\Omega)$ and $\mathbb{I}_{D1}'(\eps_p+\Omega,-\Omega)$ is the same as before with an overall minus sign for each one and therefore $\mathcal{T}_{R0}=\mathcal{T}_{0R}$.
Inserting these results into  the trace in  (\ref{trace}), the final result is given in Eqs.\,(\ref{eqzr2}) and (\ref{eqzi}). 

\noindent
\subsection{Terms quadratic in T-matrix}\label{quadterm}
The terms in the transmission probability quadratic in $G_{dd}$ are
$
\mathcal{T}^{D2}\equiv\mathcal{T}^D_{RR}+\mathcal{T}^D_{RA}+\mathcal{T}^D_{AA}+\mathcal{T}^D_{AR}
$
. It can be shown that in the DC limit the terms $\mathcal{T}^D_{RR}$ and $\mathcal{T}^D_{AA}$, which are proportional to ${(G_{dd}^R)}^2$ and ${(G_{dd}^A)}^{2}$, are zero because the propagator products involved in these terms do not produce any $\Omega^{-2}$-divergence. 
$\mathcal{T}^D_{RA}$ and $\mathcal{T}^D_{AR}$ are the complex  conjugate of each other and it suffices to calculate the first one which is
\be
\mathcal{T}^D_{RA}(\eps_p)=\frac{1}{8}\lim_{\Omega\ra 0}{I}_{D2}(\eps_p,\Omega)
{I}'_{D2}(\eps_{p},\Omega)G_{dd}^R(\eps_{p}+\Omega)G_{dd}^A(\eps_p),\label{TRA}
\ee
where the functions ${I}_{D2}$ and ${I}_{D2}'$ are defined as
\be\label{eqID2}
I_{D2}\equiv\Omega\intopi{\frac{dk_1dk_2}{(2\pi)^2}} V_{k_1}V_{k_2}\Big\{ {M}^{11}_{k_1k_2}g^A_{k_1}(\eps_p)g^R_{k_2}(\eps_{p}+\Omega)\Big\},
\ee
and
\be\label{eqID2p}
{I}'_{D2}\equiv\Omega\intopi{\frac{dq_1dq_2}{(2\pi)^2}V_{q_1}V_{q_2}}\Big\{{M}^{11}_{q_1q_2}g^R_{q_1}(\eps_{p}+\Omega)g^A_{q_2}(\eps_p)\Big\}.
\ee
Separating the contact and overlap terms of $M_{k_1k_2}^{y\psi\psi}$, defined in Eqs.\,\pref{eqdefa}-\pref{eqMy} we have
\be\label{eqMypsipsi}
M^{11}_{k_1k_2}=m_{k_1k_1}^{c}2\pi\delta(k_1-k_2)+2it\sin k_1{m^{o}_{k_1k_2}}g^R_{k_1}(\eps_{k_2}).
\ee
A factor of $2it\sin k_1$ is included for later convenience. The diagonal parameters $m_{pp}^c$ and $m_{pp}^{o}$ can be calculated from Eqs.\,\pref{eqdefa} and \pref{eqMy} are equal to
\bea\label{eqmc}
m_{pp}^c&=&\frac{\abs{\Gamma_{ep}\Gamma_{op}}}{-V_p^{2}}
\re{t_{de}^*t_{do}e^{-i(\delta^+_p-\delta^-_p)}}\br
&=&\frac{-\sqrt{1-\gamma^2}(1-\tau^2)+2\gamma\tau\sin\varphi\sin p}{1+\tau^2+2\gamma\tau\cos p\cos\varphi},
\eea
and
\bea\label{eqmo}
m_{pp}^{o}&=&\frac{-\tau\abs{\Gamma_{ep}\Gamma_{op}}^2}{\sin pV_p^2}\im{t_{de}t_{do}^*}\br
&=&\frac{-4\gamma\tau\sin\varphi\sin p}{1+\tau^2+2\gamma\tau\cos p\cos\varphi}.
\eea
Using the results of App.\,(\ref{sec:ppi}) to take the limit $\Omega \to 0$, we obtain
\bea
I_{D2}(\eps_p,\Omega)&\ra& \Big(m^{c}_{pp}+m^{o}_{pp}\Big)2\pi i\nu_p{V_p^2}
\nonumber \\
{I}'_{D2}(\eps_p,\Omega)&\ra& m^{c}_{pp}2\pi i\nu_pV_p^2.\label{eqID2p2}
\eea
Inserting these results into Eq.\,(\ref{TRA}), gives Eq.\,(\ref{eqzc2}).

\subsection{Extension to general $y$}
\label{geny}
Here we extend our results for the disconnected part of the transmission probability to the case where the source-drain voltage is applied 
asymmetrically, multiplying $N_y(t)$, defined in Eq.\,(\ref{Nydef}). This has the effect of modifying $\mathbb{A}^y_{k_1k_2}$ as indicated 
in Eq.\,(\ref{Ay}).  This has no effect on the background transmission amplitude since it gets no contribution from the contact term 
in $\mathbb{A}^y_{k_1k_2}$.  In the calculation of terms linear in the ${\rm T}$-matrix the matrices $\mathbb{I}_{D1}$ and 
$\mathbb{I}_{D1}'$ both get shifted:
\bea \mathbb{I}_{D1}&\to& \mathbb{I}_{D1}+(1-2y)V_p\mathbb{U}^\dagger_p\nonumber \\
\mathbb{I}_{D1}'&\to& \mathbb{I}_{D1}'+(1-2y)V_p\mathbb{U}_p.
\eea
In the calculation of the terms quadratic in the ${\rm T}$ the quantity $m^c_{pp}$ appearing in $M^{11}_{k_1k_2}$, 
$\mathbb{I}_{D1}$ and$ \mathbb{I}_{D1}'$  gets shifted
\be m^c_{pp}\to (2y-1)+m^c_{pp}.\ee
The background transmission probability, ${\cal T}_0(\epsilon_p)$ and $\mathcal{Z}_R$ (the coefficient of $\re{\rm T}$),  are independent of $y$. However 
\bea
&&\mathcal{Z}^y_I(\eps_p)=1-2\mathcal{T}_0(\eps_p)+(2y-1)^2\br
&&\hspace{3cm}-\frac{2\sqrt{1-\gamma^2}(1-\tau^2)(2y-1)}{1+\tau^2+2\gamma\tau\cos p\cos\varphi}\qquad\label{eqzi2y}
\eea
and
\bea
{\mathcal{Z}}^{y}_2(\eps_p)&=&\frac{-(1-\tau^2)(1-\gamma^2)+4\gamma^2\tau^2\sin^2p\sin^2\varphi}{[1+\tau^2+2\gamma\tau\cos p\cos\varphi]^2}\br
&&\hspace{.5cm}-(2y-1)^2+\frac{2\sqrt{1-\gamma^2}(1-\tau^2)(2y-1)}{1+\tau^2+2\gamma\tau\cos p\cos\varphi}\br
&=& -[2y-1+m^c_{pp}][2y-1+m^c_{pp}+m^o_{pp}]
\label{eqzc2y}
\eea
where $m^o_{pp}$ is defined in Eq.\,(\ref{eqmo}). 
Note that the sum of the two coefficients
\bea
\mathcal{Z}_{I}{'}\dn(\eps_k)&\equiv&{\mathcal{Z}}_I^{y}(\eps_k)+{\mathcal{Z}}_2^y(\eps_k)\br
&=&1-2\mathcal{T}_0(\eps_p)\br
&&\hspace{0.25cm}+\frac{4\gamma^2\tau^2\sin^2p\sin^2\varphi-(1-\tau^2)^2(1-\gamma^2)}{[1+\tau^2+2\gamma\tau\cos p\cos\varphi]^2}\br
\label{Z'def}
\eea
is independent of $y$, implying that we may write
\be\label{DeltaT}
\mathcal{T}^{yD}(\eps_p)=\mathcal{T}(\eps_p)+\Delta\mathcal{T}^{yD}
\ee
where 
\be
\Delta\mathcal{T}^{yD}=-\mathcal{Z}_2^y(\eps_p)\Delta{\rm T}_{pp}(\eps_p)
\ee
and we have separated the transmission probability into a $y$-independent part: 
\bea
\mathcal{T}(\eps_p)\equiv\mathcal{T}_0(\eps_p)&+&\mathcal{Z}_R(\eps_p)\re{ -\pi\nu_p{\rm T}_{pp}(\eps_p)}\vsp\br
&+&\mathcal{Z}_{I}{'}(\eps_p)\im{-\pi\nu_p{\rm T}_{pp}(\eps_p)}\label{Tdef}
\eea
and a part which depends on $y$ through $\mathcal{Z}_2^y(\eps_p)$. The factor multiplying this coefficient is 
\be \label{eqDeltaT}
\Delta{\rm T}_{pp}(\eps_p)\equiv\im{-\pi\nu_p{\rm T}_{pp}(\eps_p)}-\abs{-\pi\nu_p{\rm T}_{pp}(\eps_p)}^2.
\ee
The function $\Delta{\rm T}_{pp}(\eps_p)$ is equal to
deviations of the single-particle sector of the T-matrix from the optical theorem [Eq.\,\pref{eqt}] and therefore it quantifies the interaction and is 
zero for non-interacting systems.\\

\section{Non-interacting limit}
\label{sec:non-int}
In this Appendix we use Landauer and Fisher-Lee methods to calculate the conductance of the small ABK ring in the limit of a non-interacting quantum dot $U=0$ and compare the result to the one obtained from the Kubo formula. The model is the one sketeched in Fig.\,\pref{fig:Fig1} and described by the Hamiltonian \pref{eqH1}-\pref{eqH3}. 
The resulting Landauer formula for the conductance should be valid at any temperature. 
\subsection{Landauer formula}
\noindent
Starting from Hamiltonian \pref{eqH1}-\pref{eqH3} we can write down the Schr\"odinger equation $H\phi=\eps\phi$ 
and seek for a solution of the type 
\bea
\phi_n=
\left\{
\matl{
e^{ikn}+{r}e^{-ikn} & n\leq-1\\ 
{t}e^{ikn} & n\geq +1.
}
\right.
\eea
We obtain
\bw

\bea\label{eqlandauer}
{t}(\eps_k)=\frac{2i\sin k(t_Lt_Re^{-i\varphi}+t\tau(\eps_d-\eps_k))}{-t(\eps_d-\eps_k)(1-\tau^2 e^{2ik})+2t_Lt_R\tau e^{2ik}\cos\varphi+(t_L^2+t_R^2)e^{ik}}.
\eea
\ew

In order to check the consistency of this result with the Kubo calculations presented in the paper, we need to calculate the retarded Green's function of the non-interacting dot.
\subsection{Green's function of the dot}
\noindent
The Green's function of the dot is equal to
\be\label{eqgdduz}
G_{dd}^R(\eps_k)=\frac{1}{\eps_k+i\eta-\eps_d-\Sigma_{dd}^R(\eps_k)}.
\ee
At $U=0$ we can use the screening basis and Eq.\,\pref{eqvk} to write
\bea\label{eqselfdot}
\Sigma_{dd}^R(\eps_k)&=&\intopi{\frac{dq}{2\pi}\frac{V_q^2}{\eps_k-\eps_q+i\eta}}\br &=&-\frac{2t_Lt_R\tau\cos\varphi e^{2ik}+(t_L^2+t_R^2)e^{ik}}{t(1-\tau^2e^{2ik})}
\eea
where the integral is calculated using the contour technique.
In the absence of interactions, the optical theorem takes the form of Eq.\,(\ref{eqt}). 
For the dot Green's function using Eq.\,\pref{andersontmatrix} this implies
\be\label{eqgddopt}
-\pi\nu_k V_k^2\abs{G_{dd}^R(\eps_k)}^2=\im{G_{dd}^R(\eps_k)}
\ee
or in terms of the dot self-energy $-\pi\nu_kV_k^2=\im{\Sigma^R_{dd}(\eps_k)}$ 
which is indeed satisfied for the non-interacting quantum dot as can be checked from Eq.\,\pref{eqselfdot} and Eq.\,\pref{eqvk}. 

In non-interacting systems, the connected part of the transmission probability is absent ($\mathcal{T}^{yC}(\omega)=0$) and it follows from Eq.\,\pref{eqt} that the transmission probability function of Eq.\,\pref{eqTDanderson} is $y$-independent. It can be shown that by plugging in Eqs.\,\pref{eqgdduz}-\pref{eqselfdot} into the disconnected part of the transmission probability [Eq.\,\pref{eqTDanderson}] we get
\be
\mathcal{T}^D(\eps_k)=\abs{t(\eps_k)}^2.
\ee
\subsection{Fisher-Lee conductance}
An alternative approach to calculating the conductance in the non-interacting limit is to use the Fisher-Lee relation [\onlinecite{FisherLee}] which is obtained from a different version of the  Kubo formula in which $y=0$, so the voltage is 
applied to the left lead only, but the measured current is $I=dN_R/dt$. 
The connected part is absent in this non-interacting system. In this approach, the conductance has the Landauer-type form
\be\label{eqGFL1}
G\dn_{\rm FL}=\frac{2e^2}{h}\int{d\omega [-f'(\omega)]\mathcal{T}_{\rm FL}(\omega)}, \quad 
\ee
with $\mathcal{T}_{\rm FL}(\omega)=\abs{{t_{\rm FL}}(\omega)}^2$ and the transmission amplitude through the non-interacting system is given by
\be\label{eqGFL2}
{t\dn_{\rm FL}}(\eps)\equiv i\sqrt{\Gamma_L(\omega)\Gamma_R(\omega)}G_{RL}^R(\omega) \br 
\ee
where the propagator $G_{LR}^R(\omega)$ is defined by
\be
G_{RL}^R(\omega)\equiv\intoinf{dt}e^{i(\omega+i\eta) t}\braket{-i\acom{c_{1}\dn(t)}{c_{-1}\dg(0))}}.
\ee
$\Gamma_L(\eps_k)=\Gamma_R(\eps_k)=2t\sin k$ are the coupling to (or equivalently the velocity in) the leads.
In the following we use two methods to relate $G_{LR}^R(\eps)$ to the Green's function of the dot. 
\subsubsection{Keldysh approach}
The main matrix equations are those discussed in Eqs.\,\pref{eqgdl2}-\pref{eqgrl2} above. There, $L$ and $R$ refer to the first site of left and right leads, respectively. These were derived using the equation of motion technique in the Keldysh space. But here we are only interested in the equilibrium retarded Green's functions. So we take the retarded components of these equations in which $g^R_{LL}(\eps_k)=g^R_{RR}(\eps_k)=-e^{ik}/t$ is the retarded Green's function for the first site of a semi-infinite chain. From Eqs.\,\pref{eqgdl2} and \pref{eqgdr2} we get
\be
(1-\tau^2 e^{2ik})G^R_{dL}(\eps_k)=G^R_{dd}(\eps_k)\frac{e^{ik}}{t}(t_Le^{-i\varphi/2}+t_Re^{i\varphi/2}\tau {e^{ik}})
\ee
and from Eqs\,\pref{eqgll2} and \pref{eqgrl2} we have
\bea
(1-\tau^2 e^{2ik})G^R_{RL}(\eps_k)&=&-\frac{\tau e^{2ik}}{t}\br &&\hspace{-2cm}+\frac{e^{ik}}{t}(t_Re^{-i\varphi/2}
+t_Le^{i\varphi/2}\tau e^{ik})G^R_{dL}(\eps_k).\qquad\quad
\eea
$G_{LR}(\eps_k)$ is obtained from this with $L\lr R$ and $\varphi\lr-\varphi$ substitution. Combining these and using Eq.\,\pref{eqGFL2} we get
\bw

\bea
{t}\dn_{\rm FL}(\eps_k)&=&-\frac{2i\tau\sin k e^{2ik}}{1-\tau^2 e^{2ik}}
+
\frac{2ie^{2ik}\sin k(t_Le^{-i\varphi/2}+t_Re^{i\varphi/2}\tau e^{ik})(t_Re^{-i\varphi/2}+t_Le^{i\varphi/2}\tau {e^{ik}})}{t(1-\tau^2e^{2ik})^2}G^R_{dd}(\eps_k).\label{eqflk}
\eea
\ew

At $U=0$, $G^R_{dd}(\eps_k)$ is given by Eqs.\,\pref{eqgdduz}-\pref{eqselfdot}. Inserting that in Eq.\,\pref{eqflk} leads to 
\be
t\dn_{\rm FL}(\eps_k)=t(\eps_k)e^{2ik}
\ee
which is the Landauer result, Eq.\,\pref{eqlandauer} up to a factor of $e^{2ik}$ due to the propagation from site -1 to site +1.
\subsubsection{Screening and non-screening channels}
In this section we express the transmission amplitude through the ABK ring in terms of the Green's function of the dot using the $\psi$ and $\phi$ basis in the non-interacting limit.
The advantage of this method compared to the Keldysh technique is that it can be readily generalized to large rings.
We start by using Eq.\,\pref{eqqdef} to express $c\dn_{-1}$ and $c_1$ in terms of $\psi$ and $\phi$ fields 
\bea
c_{1}\dn(t)&=&\frac{1}{\sqrt{2}}\intopi{
\frac{dk}{2\pi}
\mat{\Gamma_{ek}&\Gamma_{ok}}
\mathbb{U}\dg_{k}\Psi_{k}(t)}, \br
c_{-1}\dg(0)&=&\frac{1}{\sqrt{2}}\intopi{
\frac{dq}{2\pi}
\Psi_{q}\dg(0)\mathbb{U}\dn_{q}
\mat{\Gamma^*_{eq}\\-\Gamma^*_{oq}}}.
\eea
To get the Fisher-Lee transmission amplitude $t^{\rm FL}(\eps_p)={2it\sin p} G_{RL}^R(\eps_p)$ we need to calculate the propagator
\bea
G^R_{RL}(\eps_p)
&=&\frac{1}{2}\tr{\intopi{\frac{dkdq}{(2\pi)^2}}
\mat{\Gamma_{ek}&\Gamma_{ok}}\times\br
&&\hspace{2.2cm}\mathbb{U}_{k}\dg\mathbb{G}^R_{kq}(\eps_p)\mathbb{U}\dn_{q}\mat{\Gamma^*_{eq}\\-\Gamma^*_{oq}}}.\qquad\label{eqfisher}
\eea
The Green's function matrix has the same form we encountered in Kubo calculations and can be written in terms of the T-matrix of screening and non-screening channels using Eq.\,\pref{eqgTg}. Plugging this into Eq.\,\pref{eqfisher} we get two terms which after doing the momentum integral and taking the trace can be used to write the transmission as
\bea\label{eqtfl}
t\dn_{\rm FL}(\eps_p)&=&\frac{2it\sin p}{2}\Big[\Big(\tilde{\Gamma}_{ep}-\tilde{\Gamma}_{op}\Big)\br 
&&\hspace{-1cm}+\frac{1}{2 }\Big(t_{de}^*\tilde{\Gamma}_{ep}-t_{do}^*\tilde\Gamma_{op}\Big)\Big(t_{de}\tilde\Gamma_{ep}+t_{do}\tilde\Gamma_{op}\Big)G_{dd}^R(\eps_p)\Big]\qquad.
\eea
Parameters $\tilde\Gamma_{ep}$ and $\tilde\Gamma_{op}$ containing the momentum integrals are given by
\be
\tilde{\Gamma}_{(o/e)p}\equiv\intopi{\frac{dk}{2\pi}\frac{\abs{\Gamma_{(e/o)k}}^2}{\eps_p-\eps_k+i\eta}}=\frac{e^{ip}/t}{1\pm\tau e^{ip}}
\ee
where we have used  contour integration technique and expressions for $\Gamma_{ek}$ and $\Gamma_{ok}$ are given by Eq.\,\pref{eqsmallabkgamma}. Substitution of this formula into Eq.\,\pref{eqtfl} leads to Eq.\,\pref{eqflk} obtained before. 
\subsubsection{Consistency of the Kubo conductance with the Fisher-Lee formula at $U=0$}
\noindent
Eq.\,\pref{eqflk} can be used to write
\bea
\mathcal{T}\dn_{\rm FL}(\eps_p)&\equiv&\abs{t\dn_{\rm FL}(\eps_p)}^2\vsp\br 
&=&\mathcal{T}_0^{\rm FL}(\eps_p)+\mathcal{Z}_R^{\rm FL}(\eps_p)\re{-\pi\nu_pV_p^2G_{dd}^R(\eps_p)}\vsp\br &&\hspace{.7cm}+\mathcal{Z}_I^{\rm FL}(\eps_p)\im{-\pi\nu_pV_p^2G_{dd}^R(\eps_p)}\vsp\br &&\hspace{.7cm}+\mathcal{Z}_{2}^{\rm FL}(\eps_p)\abs{-\pi\nu_pV_p^2G_{dd}^R(\eps_p)}^2\vsp
\eea
The coefficients $\mathcal{T}_0^{\rm FL}=\mathcal{T}_0$ and $\mathcal{Z}_R^{\rm FL}=\mathcal{Z}_R$ are the same as before [Eqs.\,\pref{eqz02} and \pref{eqzr2}]. 
However, $\mathcal{Z}_I^{\rm FL}$ and $\mathcal{Z}_2^{\rm FL}$ are different from the previously obtained coefficients of the T-matrix [Eqs.\,\pref{eqzi2y}, \pref{eqzc2y}] and are given by 
\bw

\bea
\mathcal{Z}_I^{\rm FL}(\eps_p)&=&-\frac{4\tau\sin p[2\tau(1+\tau^2)\sin p+\gamma(1+\tau^4)\sin\varphi+2\gamma\tau^2\sin(2p-\varphi)]}
{[(1+\tau^2)^2-4\tau^2\cos^2p](1+\tau^2+2\gamma\tau\cos p\cos\varphi)},
\\
\mathcal{Z}_2^{\rm FL}(\eps_p)&=&\frac
{4\tau^2+\gamma^2(1+\tau^4+2\tau^2\cos 2(\varphi-p))+4\tau\gamma(1+\tau^2)\cos(p-\varphi)}
{[1+\tau^2+2\gamma\tau\cos p\cos\varphi]^2}.
\eea
\ew

At $U=0$ the total conductance obtained from the two methods has to agree. In this case {the connected four-point function is absent and} the imaginary part of the single-particle T-matrix is related to its absolute value by the optical theorem [Eq.\,\pref{eqt}] and it can be checked that indeed
\be
{\mathcal{Z}}'_I(\eps_p)=
{\mathcal{Z}}_I^{\rm FL}(\eps_p)+{\mathcal{Z}}_2^{\rm FL}(\eps_p).\label{eq:FLKubo}
\ee
[${\mathcal{Z}}'_I(\eps_p)$ is defined in (\ref{Z'def}).] Comparing this to the Kubo results in Eq.\,\pref{Tdef}, we see that the two approaches give exactly the same result in the non-interacting limit.
\subsubsection{Connection with the inelastic part of the S-matrix}
The expression for the total transmission probability \pref{Tdef} is linear in the T-matrix and is valid approximately in the interacting case provided that the temperature is low enough so that the  elimination of the connected part is justified as discussed in section \pref{sec:con}. One the other hand, the Fisher-Lee conductance contains terms both linear and quadratic in the T-matrix and gives the total conductance only in the non-interacting systems. The equality \pref{eq:FLKubo} can be used to express the difference between the total conductance $G$ in Eqs.\,\pref{Tdef} and $G\dn_{\rm FL}$ as 
\bea
G-G\dn_{\rm FL}&\approx &\int{d\eps_p[-f'(\eps_p)]}\Big((\mathcal{Z}'_I-\mathcal{Z}_I^{\rm FL})\im{-\pi\nu_pV_p^2G_{dd}^R}\br
&&-\mathcal{Z}_2^{\rm FL}\abs{-\pi\nu_pV_p^2G_{dd}^R}^2\Big)\\
&=&\int{d\eps_p}[-f'(\eps_p)]{\mathcal{Z}}_2^{\rm FL}(\eps_p)\Delta{\rm T}_{pp}(\eps_p)
\eea
where we have used Eqs.\,\pref{andersontmatrix} and definition of $\Delta{\rm T}_{pp}(\eps_p)$ in Eq.\,\pref{eqDeltaT}.
This form manifestly shows that the difference $G-G\dn_{\rm FL}$ vanishes if the optical theorem for single-particle sector of T-matrix is obeyed. It is interesting to note that this difference is proportional to the inelastic part of the 
S-matrix, as defined by  Zarand et al.\,[\onlinecite{Zarand04},\onlinecite{Borda07}].

\end{document}